\documentclass[useAMS,fleqn,usenatbib]{mnras}

\usepackage{graphicx}

\usepackage{newtxtext,newtxmath}
\usepackage[T1]{fontenc}
\usepackage{ae,aecompl}
\usepackage{amsmath}	
\usepackage{amssymb}	
\usepackage{xcolor}
\usepackage{hyperref}

\title[Flux tube dynamics in accretion discs]{Dynamics of magnetic flux tubes in accretion discs of T~Tauri stars}
\author[A.~E.~Dudorov, S.~A.~Khaibrakhmanov, A.~M.~Sobolev]{A.~E.~Dudorov$^{1}$\thanks{E-mail: dudorov@csu.ru (AED)}, S.~A.~ Khaibrakhmanov$^{1,2}$\thanks{E-mail: khaibrakhmanov@csu.ru (SAKh)},
A.~M.~Sobolev$^{2}$\thanks{E-mail: andrej.sobolev@urfu.ru (AMS)},
\\
$^{1}$ Chelyabinsk state university, 129 Br. Kashirinykh str., Chelyabinsk 454001, Russia\\
$^{2}$ Ural Federal University, 51 Lenin str., Ekaterinburg 620000, Russia
}
\begin{document}

\date{}

\pagerange{\pageref{firstpage}--\pageref{lastpage}} \pubyear{2018}

\maketitle

\label{firstpage}

\begin{abstract}

Dynamics of slender magnetic flux tubes (MFT) in the accretion discs of T~Tauri
stars is investigated. We perform simulations taking into account buoyant, aerodynamic and turbulent drag forces, radiative heat exchange between MFT and ambient gas, magnetic field of the disc. The equations of MFT dynamics are solved using Runge-Kutta method of the fourth order. The simulations show that there are two regimes of MFT motion in absence of external magnetic field. In the region $r<0.2$~au, the MFT of radii $0.05 \leq a_0 \leq 0.16\,H$ ($H$ is the scale height of the disc) with initial plasma beta of 1 experience thermal oscillations above the disc. The oscillations decay over some time, and MFT continue upward motion afterwards. Thinner or thicker MFT do not oscillate. MFT velocity increases with initial radius and magnetic field strength. MFT rise periodically with velocities up to 5-15~km~s$^{-1}$ and periods of $0.5-10$~yr determined by the toroidal magnetic field generation time. Approximately 20\% of disc mass and magnetic flux can escape to disc atmosphere via the magnetic buoyancy over characteristic time of disc evolution. MFT dispersal forms expanding magnetized corona of the disc. External magnetic field causes MFT oscillations near the disc surface. These magnetic oscillations have periods from several days to 1-3 months at $r < 0.6$~au. The magnetic oscillations decay over few periods. We simulate MFT dynamics in accretion discs in the Chameleon I cluster. The simulations demonstrate that MFT oscillations can produce observed IR-variability of T Tauri stars.

\end{abstract}


\section{Introduction}

A number of observations show that young stellar objects (YSO) have large-scale magnetic field. Investigations of Zeeman splitting and broadening of spectral lines of classical T Tauri stars have shown that the stars have magnetic field with strength of $1-3$~kG at their surface~\citep{guenther99, jkrull07}. \cite{donati05} reported the registration of a magnetic field  with strength of $\sim 10^3$~G near the inner edge of the accretion disc in FU~Orionis using Zeeman splitting of spectral lines.

It is now possible to make polarization maps of the accretion discs at millimeter and (sub)millimeter wavelengths with~\textit{ALMA}. Interpretation of the polarization maps in frame of Davis-Greenstein mechanism allows to determine geometry of the magnetic field in the disc~\citep{stephens14, li16, li18}. Although there are other possible interpretations of the polarization maps~\citep[see][]{lazar07, tazaki17, kataoka17, stephens17}.

According to modern theory of star formation, young stars with the accretion discs form as a result of gravitational collapse of rotating molecular cloud cores with magnetic field~\citep[see reviews by][]{inutsuka12, li_PPVI_14}.
Numerical simulations show that the magnetic flux of the molecular cloud cores is partially conserved during the collapse, so it is natural to assume that the magnetic field of the accretion discs of young stars is the fossil one~\citep[see reviews of the theory of the fossil magnetic field by][]{dudorov95, fmft}. 
The magnetic field of accretion discs can also be a result of dynamo~\citep[see, for example,][and references therein]{brandenburg95, gressel15, moss16}.

Evolution of magnetic field in the accretion discs was usually investigated in kinematic approximation for prescribed uniform diffusivity~\citep{bkr76, lubow94, agapitou96, guilet14, okuzumi14}. 
\cite{fmfadys} and \cite{kh17} developed a magnetohydrodynamic (MHD) model of the accretion discs taking into account Ohmic, ambipolar diffusion and the Hall effect. They have shown that the magnetic field geometry varies through the disc. Inside the region of low ionization fraction (`dead' zone~\citep{gammie96}), Ohmic diffusion hinders amplification of the magnetic field, so that the magnetic field has poloidal geometry. Ambipolar diffusion operates in outer regions of the accretion discs, where the magnetic field asquires quasi-radial or quasi-azimuthal geometry depending on the intensity of ionization and grain parameters. The Hall effect operates near the borders of the `dead' zones and leads to redistribution of the poloidal and toroidal components of the magnetic field. The magnetic field is frozen in gas near the inner edge of the accretion discs, where thermal ionization operates. Effective generation of the toroidal magnetic field is possible in this region.  \cite{fmfadys} have suggested that magnetic buoyancy instability can solve the problem of runaway growth of toroidal magnetic field.

The magnetic buoyancy instability leads to formation of separate magnetic flux tubes (MFT) from the regular magnetic field~\citep[see][]{parker_book}. The MFT can float from the region of their formation to the surface under the action of buoyant force. The instability has been found in numerical simulations of unstable gas layers with planar magnetic field \citep{cattaneo88, matthews95, wissink00, fan01}. \cite{vasil08} have shown that the instability also arises in the gas layer with the magnetic field that is generated out of perpendicular magnetic field by a shear flow in the plane of the layer. \cite{takasao18} detected formation of MFT in 3D MHD simulations of the inner regions of the magnetized accretion discs of young stars.

Magnetic buoyancy instability develops and MFT form if the magnetic pressure is of the order of the gas pressure. Nonlinear evolution of the instability can lead to formation of the MFT with strong magnetic field (plasma $\beta<1$). For example, \cite{machida00} performed MHD simulations of magnetic buoyancy instability in differentially rotating magnetized disks. They reported about the formation of a filamentary-shaped  intermittent magnetic structures ($\beta<1$) inside the disc. Inside the disc, magnetic buoyancy instability will lead to formation of magnetic rings of the toroidal magnetic field. Major radius of the rings will be equal to the distance to the star, while minor radius will be limited by the pressure scale height of the disc. 

The MFT rise upwards  in the disc under the action of buoyant force, because their density is less than the density of ambient gas. The MFT dynamics in the discs has been investigated numerically in frame of slender magnetic flux tube approximation \citep{sakimoto89, torkelsson93, chakra94, schram96, achterberg96a}.  
\cite{sakimoto89} investigated the dynamics of MFT inside the radiation pressure dominated regions of accretion discs of quasi-stellar objects taking into account aerodynamic drag, MFT shear, and heat exchange between the MFT and external gas in radiative diffusion approximation. \cite{torkelsson93} considered similar problem for the case of Stokes' drag law. \cite{schram96} studied MFT dynamics in optically thick radiation pressure dominated discs paying special attention to the role of shear and magnetic tension. He have found that strong shear can lead to formation of coronal loops from the initially horizontal toroidal MFT.
\cite{chakra94} investigated dynamics of toroidal MFT inside the geometrically thick radiation pressure dominated discs around black holes. They have shown that magnetic tension leads to MFT collapse.  \cite{deb17} extended model of \cite{chakra94} to take into account time-dependent evolution of the flow around the MFT. 

Dynamics of slender MFT in  accretion discs of young stars, and effect of external magnetic field on the dynamics of the MFT  have not been investigated yet. 
Some of the results of simulations in slender flux tube approximation were confirmed in 2D and 3D MHD simulations of the magnetic flux escape from the accretion discs~\citep{matsumoto88, shibata90, ziegler01} and stratified gas layer~\citep{sykora15}.

After rising from the disc, the MFT can lead to various effects, such as outflows, variability and bursts. \cite{dudorov91rings} proposed that MFT can be a part of the molecular outflows in the star formation regions. \cite{chakra94} and \cite{deb17} argued that the MFT can play a role in acceleration and collimation of jets from accretion discs of black holes. 
Formation of looplike magnetic structures above the discs due to magnetic buoyancy instability and magnetic reconnection can lead to burst acitivity and heating of the region above the disc~\citep{galeev79, stella84, schram96, miller00, hirose11, uzdensky13}. \cite{mazets93} proposed that the MFT can carry away angular momentum from the accretion discs. 

Present paper concerns mass and magnetic flux escape from the accretion discs of young stars due to magnetic buoyancy, formation of expanding magnetized `corona' above the disc, and connection between rising MFT and IR-variability of the accretion discs. We investigate MFT dynamics in the accretion discs of young stars taking into account the radiative heat exchange between MFT and ambient gas, effects of the aerodynamic and turbulent drag, and external magnetic field. Some particular aspects of MFT  dynamics in the accretion discs of young stars were considered by~\cite{bmfad} and~\cite{kh17raa}. Structure of the accretion disc is calculated with the help our MHD model of the accretion discs~\citep{fmfadys, kh17}.

The paper is organized as follows. In section~\ref{Sec:problem}, we discuss approximations of the model. In section~\ref{Sec:eqs}, the governing equations are derived. The governing equations are written in terms of non-dimensional variables in section~\ref{Sec:nondim}. Section~\ref{Sec:methods} is devoted to solution methods of the model equations. The model of the disc is described in section~\ref{Sec:disk}. Fiducial results are presented and discussed in section~\ref{Sec:Fiduc}. In section~\ref{Sec:Param}, we investigate influence of the model parameters on the MFT dynamics. We estimate mass loss rates due to rising MFT in section~\ref{Sec:outflows}. The MFT dynamics taking into account magnetic pressure of the disc is investigated in section~\ref{Sec:oscill}. We apply our model for interpretation of the observational data on IR-variability in section~\ref{Sec:observ}. Section~\ref{Sec:end} summarize results and conclusions.

\section{Model}

\subsection{Problem statement}
\label{Sec:problem}

\begin{figure*}
\begin{center}
\includegraphics[width=0.99\textwidth]{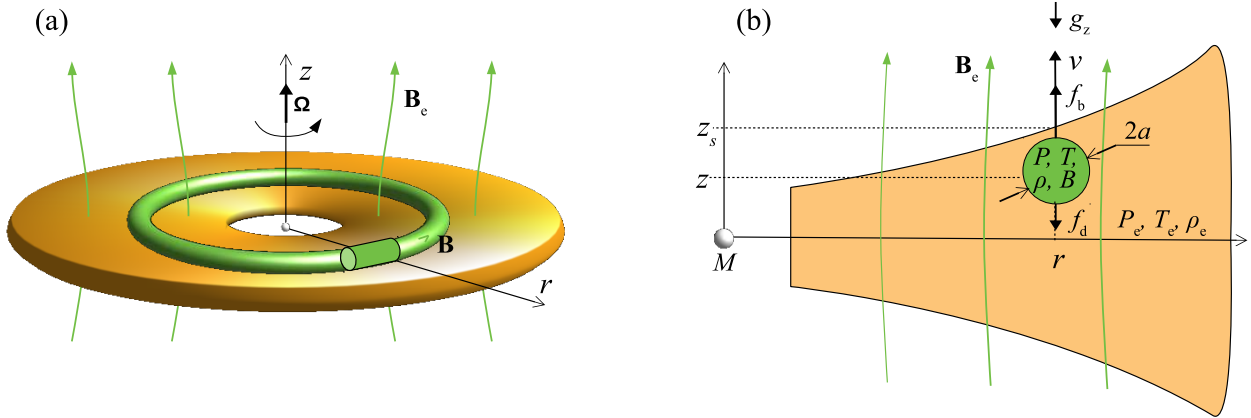}
\end{center}
\caption{Panel (a): general picture of the accretion disc (orange color) with magnetic field $B_{{\rm e}}$ and toroidal magnetic flux tube with $B$ (green color). Panel (b): cross-section of the disc and magnetic flux tube in the $r-z$ plane. Dynamics of slender cylindrical MFT in the $z$-direction under the action of buoyant force, $f_{{\rm b}}$, and drag force, $f_{{\rm d}}$, is investigated. (color figure online)}
\label{Fig:scheme}
\end{figure*}

We consider geometrically thin, optically thick accretion disc of a young star (see Figure~\ref{Fig:scheme}). The disc has pressure $P_{{\rm e}}$, density $\rho_{{\rm e}}$, temperature $T_{{\rm e}}$, and magnetic field with strength $B_{{\rm e}}$. Disc mass is small compared to the mass of the star $M_{\star}$, and therefore self-gravity of the disc can be neglected. We use the cylindrical system of coordinates $(r,\, \varphi,\, z)$. The vertical axis $z$ is directed along the angular velocity vector of the disc ${\bf \Omega}=(0,\,0,\,\Omega)$. The accretion disc is considered to be in hydrostatic equilibrium in the $z$-direction. Vertical coordinate of the surface of the disc is $z_{{\rm s}}$.

We assume that magnetic buoyancy instability leads to formation of a MFT in the form of a torus out of toroidal magnetic field in the disc.  In the case of axial symmetry, we investigate the dynamics of the unit length cylindrical element of this magnetic torus. The MFT is located at the distance $r$ from the rotation axis at a coordinate $z_0$ in the beginning.  It has velocity ${\bf v}$, cross-section radius $a$, pressure $P$, density $\rho$, temperature $T$, and magnetic field strength $B$. Schematic problem statement is shown in Figure~\ref{Fig:scheme}.

The MFT is in pressure equilibrium with the external gas,
\begin{equation}
P + \frac{B^2}{8\upi} = P_{{\rm e}}.\nonumber
\end{equation}
Magnetic pressure outside the MFT is not considered in this equation. The effect of external magnetic field is considered in Section~\ref{Sec:oscill}. We assume that temperatures inside and outside the MFT are equal to each other initially.   For the ideal gas with equation of state
\begin{equation}
	P = \frac{R_{{\rm g}}}{\umu}\rho T\label{Eq:eos}\\
\end{equation}
the density difference equals
\begin{equation}
	\Delta \rho=\rho_{{\rm e}}-\rho = \frac{B^2}{8\upi v_{{\rm s}}^2},\label{Eq:drho}
\end{equation}
where 
\begin{equation}
	v_{{\rm s}} = \sqrt{\frac{R_{{\rm g}}T_{{\rm e}}}{\umu}}
\end{equation}
is the sound speed, $R_{{\rm g}}$ is the universal gas constant, $\umu=2.3$ is the mean molecular weight of the gas. The gas inside the MFT has smaller density comparing to the surrounding gas, as $P<P_{{\rm e}}$. Due to positive density difference $\Delta\rho>0$, the buoyant force 
\begin{equation}
	f_{{\rm b}} = -\Delta\rho g_z,\label{Eq:f_b}
\end{equation}
causes rise of the MFT in the $z$-direction, where $g_z<0$ is the vertical component of stellar gravity. Drag force counteracts the motion of the MFT. We consider drag forces of two types, turbulent and aerodynamic drag.

\subsection{Basic equations}
\label{Sec:eqs}

Our model of MFT dynamics is based on the slender flux tube approximation. 
Similar models have been used by~\cite{sakimoto89}, \cite{torkelsson93}, \cite{chakra94}, \cite{schram96}. We write the equations of the MFT dynamics following~\cite{dk85},
\begin{eqnarray}
	\frac{{\rm d}{\bf v}}{{\rm d}t} &=& \left(1 - \frac{\rho_{{\rm e}}}{\rho}\right)\mathbfit{ g} + \mathbfit{ f}_{{\rm d}}\left(\mathbfit{ v},\, \rho,\, T,\, a,\, \rho_{{\rm e}}\right),\label{Eq:motion}\\
	\frac{{\rm d}{\bf r}}{{\rm d}t} &=& \mathbfit{ v},\label{Eq:velocity}\\
	M_{{\rm l}} &=& \rho\upi a^2 = {\rm const},\label{Eq:mass}\\
	\Phi &=& \upi a^2B ={\rm const},\label{Eq:mflux}\\
	{\rm d}Q &=& {\rm d}U + P_{{\rm e}}{\rm d}V,\label{Eq:dQ}\\
	P + \frac{B^2}{8\upi} &=& P_{{\rm e}},\label{Eq:pbal2}\\
	\frac{{\rm d}P_{{\rm e}}}{{\rm d}z} &=& -\rho_{{\rm e}}g_z,\label{Eq:disc}\\
	U &=& \frac{P_{{\rm e}}}{\rho(\gamma - 1)} + \frac{B^2}{8\upi\rho}.\label{Eq:eos_kalor}
\end{eqnarray}
In equation of motion (\ref{Eq:motion}) $\mathbfit{ f}_{{\rm d}}$ is the drag force per unit mass of the flux tube, (\ref{Eq:velocity}, \ref{Eq:mass}, \ref{Eq:mflux}) are the equations defining the velocity $\mathbfit{ v}$, mass $M_{{\rm l}}$ per unit length and magnetic flux $\Phi$ of the MFT, (\ref{Eq:dQ}) is the first law of thermodynamics ($Q$ is the quantity of heat per unit mass, $U$ is the energy of MFT per unit mass, $V=1/\rho$ is the specific volume), (\ref{Eq:pbal2}) is the pressure balance equation, (\ref{Eq:disc}) is the equation of the hydrostatic equilibrium of disc in the $z$-direction, (\ref{Eq:eos}) is the equation of state, $\gamma$ is the adiabatic index.

We consider MFT motion in the $z$-direction, $\mathbfit{ v} = \left(0,\, 0,\, v\right)$, $\mathbfit{ f}_{{\rm d}} = \left(0,\, 0,\, f_{{\rm d}}\right)$. In this case, Equations (\ref{Eq:motion}, \ref{Eq:velocity}) are reduced to
\begin{eqnarray}
	\frac{{\rm d}v}{{\rm d}t} &=& \left(1 - \frac{\rho_{{\rm e}}}{\rho}\right)g_z + f_{{\rm d}},\label{Eq:v}\\
	\frac{{\rm d}z}{{\rm d}t} &=& v.\label{Eq:z}
\end{eqnarray}

The vertical component of stellar gravity acceleration
\begin{equation}
	g_z = -z\frac{GM_{\star}}{r^3}\left(1 + \frac{z^2}{r^2}\right)^{-3/2}.\label{Eq:gz}
\end{equation}

In the case of aerodynamic drag (see~\cite{parker_book})
\begin{equation}
	f_{{\rm d}} = -\frac{\rho_{{\rm e}}v^2}{2}\frac{C_{{\rm d}}}{\rho\upi a},\label{Eq:fa}
\end{equation}
where $C_{{\rm d}}$ is the drag coefficient $\sim 1$. The turbulent drag force can be calculated as~\citep{pneuman72}
\begin{equation}
	f_{{\rm d}} = -\frac{\upi\rho_{{\rm e}}\left(\nu_{{\rm t}}av^3\right)^{1/2}}{\rho\upi a^2},\label{Eq:ft}
\end{equation}
Turbulent viscosity $\nu_{{\rm t}}$ can be estimated as \citep{shakura72, ss73}
\begin{equation}
	\nu_{{\rm t}} = \alpha v_{{\rm s}} H,\label{Eq:nu_t}
\end{equation}
where $\alpha$ is non-dimensional parameter characterizing the turbulence efficiency, $H$ is the scale height of the disc. Drag force is evaluated using formula~(\ref{Eq:ft}) inside the disc, $z\leq z_{{\rm s}}$, and using formula~(\ref{Eq:fa}) above the disc, $z> z_{{\rm s}}$.

Equations (\ref{Eq:mass}) and (\ref{Eq:mflux}) give
\begin{eqnarray}
	a &=& a_0\left(\frac{\rho}{\rho_0}\right)^{-1/2},\label{Eq:a}\\
	B &=& B_0\frac{\rho}{\rho_0},\label{Eq:B}
\end{eqnarray}
where $a_0$, $\rho_0$ and $B_0$ are the initial radius, density and magnetic field strength of the MFT, respectively.

Taking time derivative of Equations (\ref{Eq:dQ}) and (\ref{Eq:pbal2}) we obtain
\begin{eqnarray}
	\frac{{\rm d}U}{{\rm d}t} - \frac{P_{{\rm e}}}{\rho^2}\frac{{\rm d}\rho}{{\rm d}t} &=& h_{{\rm c}},\label{Eq:hc}\\
	\frac{{\rm d}}{{\rm d}t}\left(P + \frac{B^2}{8\upi}\right) &=& v\frac{{\rm d}P_{{\rm e}}}{{\rm d}z},\label{Eq:dPdt}
\end{eqnarray}
where
\begin{equation}
	h_{{\rm c}} = \frac{{\rm d}Q}{{\rm d}t}
\end{equation}
is the rate of heat exchange. Energy and pressure are the thermodynamic functions of density and temperature, $U=U\left(\rho,\,T\right)$ and $P = P(\rho,\, T)$. Then equations (\ref{Eq:hc}) and (\ref{Eq:dPdt}) can be written as
\begin{eqnarray}
	U_T\frac{{\rm d}T}{{\rm d}t} + \left(U_{\rho} - \frac{P_{{\rm e}}}{\rho^2}\right)\frac{{\rm d}\rho}{{\rm d}t} &=& h_{{\rm c}},\label{Eq:hc2}\\
	\left(P_{\rho} + C_{{\rm m}}\rho\right)\frac{{\rm d}\rho}{{\rm d}t} + P_{T}\frac{{\rm d}T}{{\rm d}t} &=& v\frac{{\rm d}P_{{\rm e}}}{{\rm d}z},\label{Eq:dpe_dt}
\end{eqnarray}
where the superscript $\left(...\right)_T$ means derivative with respect to $T$ (with constant $\rho$) and the superscript $\left(...\right)_{\rho}$ means derivative with respect to $\rho$ (with constant $T$). Term $C_{{\rm m}}\rho$ is the derivative of magnetic pressure $B^2/8\upi$ with respect to $\rho$, and $C_{{\rm m}}=\dfrac{B_0^2}{4\upi\rho_0^2}$.
 
Solving equations (\ref{Eq:hc2}, \ref{Eq:dpe_dt}) for time derivatives of $\rho$ and $T$ and using (\ref{Eq:disc}), we derive equations describing the evolution of density and temperature of the MFT
\begin{eqnarray}
	\frac{{\rm d}\rho}{{\rm d}t} &=& \frac{h_{{\rm c}}P_T + U_T\rho_{{\rm e}} g_z v}{P_T\left(U_\rho - \dfrac{P_{{\rm e}}}{\rho^2}\right) - U_T\left(P_\rho + C_{{\rm m}}\rho\right)},\label{Eq:rho}\\
	\frac{{\rm d}T}{{\rm d}t} &=& \frac{\rho_{{\rm e}} g_z v\left(U_\rho - \dfrac{P_{{\rm e}}}{\rho^2}\right) + h_{{\rm c}}\left(P_\rho + C_{{\rm m}}\rho\right)}{U_T\left(P_\rho + C_{{\rm m}}\rho\right) - P_T\left(U_\rho - \dfrac{P_{{\rm e}}}{\rho^2}\right)}.\label{Eq:T}
\end{eqnarray}

Now consider equations of the accretion disc structure. We solve equation of the hydrostatic equilibrium (\ref{Eq:disc}) using polytropic dependence of pressure on density together with equation of state (\ref{Eq:eos}), and get the vertical profiles of density and temperature inside the disc
\begin{eqnarray}
	\rho_{{\rm e}}(z) &=& \rho_{{\rm m}}\left(1 - \frac{k-1}{2k}\left(\frac{z}{H}\right)^2\right)^{\frac{1}{k-1}},\label{Eq:rhoe}\\
	T_{{\rm e}}(z) &=& T_{{\rm m}}\left(1 - \frac{k-1}{2k}\left(\frac{z}{H}\right)^2\right),\label{Eq:Te}
\end{eqnarray}
where $\rho_{{\rm m}}=\rho_{{\rm e}}(z=0)$, $T_{{\rm m}}=T_{{\rm e}}(z=0)$ are the density and temperature in the midplane of the disc, $k=1+1/n$, $n$ is the polytropic index, scale height $H=v_{{\rm s}}/\Omega_{{\rm k}}$,
\begin{equation}
	\Omega_{{\rm k}} = \sqrt{\frac{GM_{\star}}{r^3}}
\end{equation}
is the Keplerian angular velocity.

We set temperature $T_{{\rm a}}$ of the gas above the disc, $z>z_{{\rm s}}$, equal to the effective temperature of the disc
\begin{equation}
	T_{{\rm a}} = T_{{\rm eff}} = 280\,\left(\frac{L}{L_{\sun}}\right)^{1/4}\left(\frac{r}{1\,{\rm au}}\right)^{-1/2}\,{\rm K},\label{Eq:Tatm}
\end{equation} 
where $L$ is the luminosity of the star. Formula (\ref{Eq:Tatm}) is derived under the assumption that the gas in the photosphere of the disc is heated by stellar radiation~\citep{hayashi81}. We determine the coordinate of the disc surface from the equality of (\ref{Eq:Te}) and effective temperature of the disc (\ref{Eq:Tatm}) 
\begin{equation}
	z_{{\rm s}} = H\sqrt{\frac{2k}{k-1}\left(1 - \frac{T_{{\rm a}}}{T_{{\rm m}}}\right)}.
\end{equation}

We assume that  temperature is constant $T_{{\rm a}}$ above the disc surface, $z> z_{{\rm s}}$, and density falls down with $z$ according to Equation (\ref{Eq:disc}) to the point where $\rho_{{\rm e}}$ becomes equal to the density of molecular cloud core $\rho_{{\rm ism}} = 3.8 \times 10^{-20}\,{\rm g}\,{\rm cm}^{-3}$.

Heating rate of the MFT can be evaluated from heat transfer equation~\citep[see][]{zeldovich_raizer}
\begin{equation}
	h_{{\rm c}} = \frac{1}{\rho}{\rm div}\mathbfit{ q},
\end{equation}
where $\mathbfit{ q}$ is the vector of the heat flux density. We consider the heat flux driven by the radiative heat conductivity~\citep{mihalas_book}
\begin{equation}
	\mathbfit{ q} = -\kappa\nabla T.
\end{equation}
\begin{equation}
	\kappa = \frac{4\sigma_{{\rm R}} T^3}{3\kappa_{{\rm R}}\rho},
\end{equation}
where $\kappa_{{\rm R}}$ is the Rosseland mean opacity, $\sigma_{{\rm R}}$ is the Stefan-Boltzmann constant. The heat exchange occurs through the surface of the MFT. Let us introduce cylindrical coordinates $(r',\, \varphi',\, z')$, where $r'$ is the distance from the axis of the MFT, $\varphi'$ is the azimuthal angle, $z'$ is the coordinate along the axis of the MFT. Then
\begin{equation}
{\rm div}{\mathbfit q} = \frac{1}{r'}\frac{\upartial}{\upartial r'} \left(r'q\right)\approx \frac{q_{{\rm ext}} - q_{{\rm in}}}{a},
\end{equation}
where $q_{{\rm ext}}$ and $q_{{\rm in}}$ are external and internal heat flux densities, correspondingly. Then
\begin{equation}
	h_{{\rm c}} \simeq -\frac{4}{3\kappa_{{\rm R}}\rho^2}\frac{\sigma_{{\rm R}}T^4 - \sigma_{{\rm R}}T_{{\rm e}}^4}{a^2}.
\end{equation}
We determine $\kappa_{{\rm R}}$ as the power-law function of gas density and temperature following~\citet{fmfadys}. 

\subsection{Non-dimensional variables}
\label{Sec:nondim}

Let us introduce following non-dimensional variables:
\begin{eqnarray}
	u = v/v_{{\rm a}}, & \tilde{z} = z / H, & \tilde{T} = T / T_{{\rm m}}, \nonumber\\ \tilde{\rho} = \rho / \rho_{{\rm m}}, & \tilde{t} = t / t_{{\rm A}}, & \tilde{h}_{{\rm c}} = h_{{\rm c}} / h_{{\rm m}},\nonumber\\
	 \tilde{B} = B / B_{{\rm e}}, &	\tilde{g} = g_z/f_{{\rm a}}, & \tilde{f}_{{\rm d}} = f_{{\rm d}} / f_{{\rm a}} , \nonumber\\
	 & \tilde{P} = P / (\rho_m v_{{\rm a}}^2), & 
\end{eqnarray}
where $v_{{\rm a}}$ is the Alfv{\'e}n speed, $t_{{\rm A}} = H / v_{{\rm a}}$ is the Alfv{\'e}n crossing time,  $h_{{\rm m}}={\varepsilon}_{{\rm m}}/t_{{\rm A}}$, ${\varepsilon}_{{\rm m}}$ is the energy density of magnetic field, $B_{{\rm e}}$ is the magnetic field strength, $f_{{\rm a}} = v_{{\rm a}} / t_{{\rm A}}$. All scales are defined at the midplane of the disc.

Equations (\ref{Eq:v}, \ref{Eq:z}, \ref{Eq:rho}, \ref{Eq:T}, \ref{Eq:a}, \ref{Eq:B}) in the non-dimensional variables (tilde signs are omitted):
\begin{eqnarray}
	\frac{{\rm d}u}{{\rm d}t} &=& \left(1 - \frac{\rho_{{\rm e}}}{\rho}\right)g + f_{{\rm d}},\label{Eq:u}\\
	\frac{{\rm d}z}{{\rm d}t} &=& u,\label{Eq:xi}\\
		\frac{{\rm d}T}{{\rm d}t} &=& \frac{2\left(\gamma - 1\right)}{\beta}\times \nonumber\\
	& & \frac{h_{{\rm c}}\left(\dfrac{\beta}{2}T + C_{{\rm m}}\rho\right) + \rho_e gu\left(\dfrac{C_{{\rm m}}}{2} - \dfrac{P_{{\rm e}}}{\rho}\right)}{\dfrac{3-\gamma}{2}C_{{\rm m}}\rho + \dfrac{\beta}{2}T + \left(\gamma - 1\right)\dfrac{P_{{\rm e}}}{\rho}},\label{Eq:tT}\\
	\frac{{\rm d}\rho}{{\rm d}t} &=& -\frac{\rho_{{\rm e}} gu + (\gamma - 1)h_{{\rm c}}\rho}{\dfrac{3-\gamma}{2}C_{{\rm m}}\rho + \dfrac{\beta}{2}T + \left(\gamma - 1\right)\dfrac{P_{{\rm e}}}{\rho}},\label{Eq:trho}\\
	a &=& C_{{\rm a}} \rho^{-1/2},\label{Eq:ta}\\
	B &=& C_{{\rm B}} \rho,\label{Eq:tB}
\end{eqnarray}
where $\beta$ is the plasma beta calculated for the parameters of the disc in the midplane,
\begin{equation}
	C_{{\rm B}} = \frac{\tilde{B}_0}{\tilde{\rho}_0},
\end{equation}
\begin{equation}
	C_{{\rm a}} = \tilde{a}_0\tilde{\rho}_0^{1/2}.
\end{equation}

\subsection{Method of solution and initial conditions}
\label{Sec:methods}\

The system of dynamic equations (\ref{Eq:u}-\ref{Eq:trho}) is solved with the help of the explicit Runge-Kutta method of the fourth order of accuracy. Automatic step selection is used, and adopted relative accuracy equals $10^{-4}$. At each time step, the radius and magnetic field strength of the MFT are calculated from Equations (\ref{Eq:ta}-\ref{Eq:tB}).

At the initial moment of time, the MFT with coordinates $r$ and $z_0$ has zero velocity $u_0=0$. The MFT is in thermal equilibrium with surrounding gas, $T(z_0)=T_{{\rm e}}$. We specify the initial magnetic field strength of the MFT $B_0$ with the help of plasma beta
\begin{equation}
	B_0= \sqrt{\frac{8\upi P_{0}}{\beta_0}},\label{Eq:beta0}
\end{equation}
where $\beta_{0}$ is the initial plasma beta inside the MFT. The initial density is determined from the condition of the pressure equilibrium (\ref{Eq:pbal2}) using plasma beta,
\begin{equation}
\rho_0 = \dfrac{P_{{\rm e}}(z_0)}{\dfrac{R_gT(z_0)}{\umu}\left(1 + \dfrac{1}{\beta_0}\right)}.
\end{equation}

\subsection{Model of the disc}
\label{Sec:disk}
We use our MHD model of the accretion discs to calculate the structure and magnetic field of the accretion disc. Let us describe briefly the features of the model (see~\cite{fmfadys} and \cite{kh17} for details).

The model is MHD-generalization of~\cite{ss73} model. We solve MHD equations in the approximation of a geometrically thin and optically thick stationary disc. It is assumed that the turbulence is the main mechanism of the angular momentum transport in the disc. Turbulent viscosity is estimated according to expression~(\ref{Eq:nu_t}). 
The temperature of the disc is calculated from the balance between turbulent `viscous' heating and radiative cooling following~\cite{ss73}. We use low-temperature opacities from~\cite{semenov03}. The heating of the outer parts of the disc by stellar radiation and cosmic rays is also taken into account following~\cite{dalessio98}. These two mechanisms determine the temperature of the disc in the regions, where the turbulence can be weak. The model has two main parameters: $\alpha$ and accretion rate~$\dot{M}$.

In addition to equations of \cite{ss73}, we solve the induction equation taking into account Ohmic diffusion, magnetic ambipolar diffusion, magnetic buoyancy and the Hall effect. Ionization fraction is determined from the equation of collisional ionization \citep[see][]{spitzer_book} considering the ionization by cosmic rays, X-rays and radioactive decay, radiative recombinations and the recombinations on dust grains. The evaporation of dust grains and thermal ionization of hydrogen and metals are included in the model following~\cite{dud87}.

Inner radius of the disc is assumed to be equal to the radius of stellar magnetosphere. Outer radius of the disc, $r_{\rm{out}}$, is determined as the contact boundary, where the disc pressure equals the pressure of the external medium.

\section{Results}

We consider accretion disc of classical T Tauri star with mass $M_{\star}=1\,M_{\sun}$, radius $R_{\star}=2\,R_{\odot}$, surface magnetic field strength $B_{\star}=2$~kG, liminocity $L_{\star}=1\,L_{\odot}$. The disc is characterized by turbulence parameter $\alpha=0.01$ and mass accretion rate $\dot{M}=10^{-7}\,M_{\sun}\,{\rm yr}^{-1}$. In this case, the inner radius of the disc lies at $r_{{\rm in}}=0.027$~au from the star, and the outer radius $r_{{\rm out}}=320$~au. Ionization fraction $x$ in the disc is calculated for dust grain radius $0.1\,\mu$m, cosmic rays ionization rate $\xi=10^{-17}\,{\rm s}^{-1}$ and attenuation length $\Sigma_{{\rm CR}}=100\,{\rm g}\,{\rm cm}^{-2}$. 

\begin{figure}
\begin{center}
\includegraphics[width=0.49\textwidth]{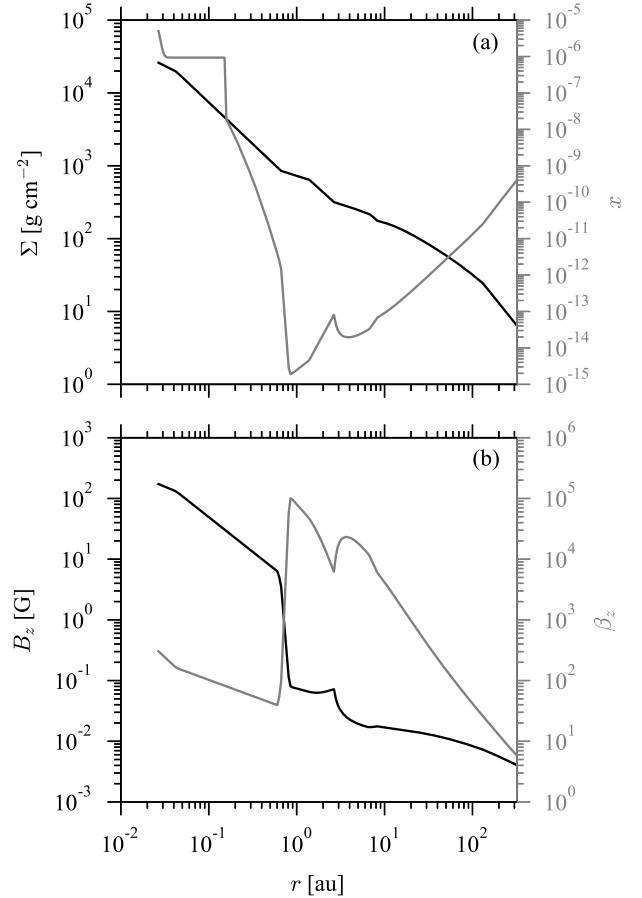}
\end{center}
\caption{Panel (a): radial profiles of the surface density, $\Sigma$ (black line, left $y$-axis), and ionization fraction, $x$ (grey line, right $y$-axis), in the MHD model of the accretion disk for the adopted parameters. Panel (b): radial profiles of $B_z$ (black line, left $y$-axis) and corresponding plasma beta (grey line, right $y$-axis).}
\label{Fig:beta}
\end{figure}

In Figure~\ref{Fig:beta}a, we plot radial profiles of gas surface density and ionization fraction of the disc calculated using our model for adopted parameters. Surface density decreases with $r$ from $\approx 2.5\times 10^4\,\mathrm{g}\,\mathrm{cm}^{-2}$ at the inner boundary of the disc to $\approx 5\,\mathrm{g}\,\mathrm{cm}^{-2}$ near its outer boundary. Typical slope of $\Sigma(r)$ dependence is $-0.7$ in the region $r=[1,\,100]$~au. The radial profile of ionization fraction is non-monotonic. In the innermost part of the disk, $r < 0.6$ au, the ionization fraction is high, $x > 10^{-10}$, due to thermal ionization. A plateau in $x(r)$ profile at $r=[0.03,\,0.2]$~au reflects total ionization of Potassium. The ionization fraction is minimum at $r\approx 0.8$~au. At further distances, ionization fraction increases with $r$ due the decrease of surface density and corresponding more efficient ionization  by cosmic rays.

In Figure~\ref{Fig:beta}b, we plot radial profiles of $B_z$ and corresponding plasma beta calculated as $\beta_z = 8\pi \rho_{{\rm m}} v_{{\rm s}}^2 / B_z^2$. Magnetic field strength decreases with distance $r$. Near the inner edge of the disk, $B_z\approx 170$~G, which is nearly equal to the stellar magnetic field at this distance. In the region of thermal ionization, $r=[0.027,\, 0.6]$~au, the magnetic field is frozen into gas, and the strength of $B_z$ is proportional to gas surface density. Plasma beta decreases from 300 to 30 in this region. Magnetic field strength abruptly decreases by two orders of magnitude at $r\simeq 0.6$~au. This is the transition to the `dead' zone, region of low ionization fraction where Ohmic and ambipolar diffusion prevent magnetic field amplification. The `dead' zone occupies region from 0.6 to 30 au for considered parameters. Plasma beta is $10^3-10^5$ inside the `dead' zone. Further $B_z$ and $\beta_z$ decrease with $r$ to the values $5\times 10^{-3}$~G and $6$, respectively, near the outer edge of the disc.

In the following we consider dynamics of the MFT that can form in the inner region of the disc, $r<0.6$~au. We carry out simulations of the MFT dynamics for the various initial radii $a_0=\left[0.01,\, 0.1,\, 0.2,\, 0.4\right]\,H$, plasma betas $\beta_0 = [0.01$,  $0.1$, $1$, $10]$, coordinates $r$ in range $0.027 \div 0.6$~au and coordinates $z_0=[0.5,\,1]\,H$. Adiabatic index of the gas $\gamma = 7/5$.

Density, temperature and magnetic field strength of the accretion disc are listed in Table~\ref{Tab:A} (column~1: $r$-distance, column~2: midplane density, column~3: midplane temperature, column~4: effective temperature of the disc, column~5: scale height, column~6: magnetic field strength $B_{{\rm e}} = B_z$ in the disc, column~7: $z$-coordinate of the surface of the disc).  Table~\ref{Tab:A} shows that density, temperature and magnetic field strength decrease with $r$-distance in the disc. Scale height of the disc increases with $r$. We choose polytropic index $n=3$. In this case, the coordinate of the disc surface varies from $z_{{\rm s}}=2.28\,H\approx 0.0014$~au at $r=0.027$~au to $z_{{\rm s}}=2.38\,H\approx 0.08$~au at $r=0.6$~au.

\subsection{Fiducial run}
\label{Sec:Fiduc}
In this section, we present and discuss the simulations of MFT dynamics for the following representative parameters: $r=0.15$~au, $\beta_0=1$, $a_0=0.1\,H$, and $z_0=0.5\,H$. We perform two sets of simulations to study the role of radiative heat exchange. First, we simulate the dynamics of MFT evolving in thermal equilibrium with ambient gas (Section~\ref{Sec:thb}). In this case, Equations~(\ref{Eq:tT}) and (\ref{Eq:trho}) are excluded, equality $T=T_{{\rm e}}$ is adopted, and density is determined from pressure balance Equation~(\ref{Eq:pbal2}). Second, we carry out the simulations taking into account the radiative heat exchange according to Equations~(\ref{Eq:tT}) and (\ref{Eq:trho}) (Section~\ref{Sec:he}).

\begin{table}
\caption{The characteristics of the accretion disc}
\centering
\begin{tabular}{@{}cccccc}
\hline
$r$ [au] & $\rho_{{\rm m}}$ [g cm$^{-3}$] & $T_{{\rm m}}$ [K] & $T_{{\rm a}}$ [K] & $H$ [au] & $B_{{\rm e}}$ [G] \\ 
(1) & (2) & (3) & (4) & (5) & (6)  \\
\hline 
\\
0.027 & $2.0\times 10^{-6}$ & $4830$ & $1700$ & $6.2\times 10^{-4}$ & $170$ \\ 
0.15 & $4.1\times 10^{-8}$ & $2015$ & $715$ & $5.3\times 10^{-3}$ & $29.5$ \\ 
0.2 & $2.1\times 10^{-8}$ & $1840$ & $625$ & $7.7\times 10^{-3}$ & $22.4$ \\ 
0.4 & $3.7\times 10^{-9}$ & $1430$ & $445$ & $2.0\times 10^{-2}$ & $9.9$ \\ 
0.6 & $1.4\times 10^{-9}$ & $1240$ & $360$ & $3.3\times 10^{-2}$ & $6.3$ \\ 
\\
\hline 
\end{tabular} 
\label{Tab:A}
\end{table}

\subsubsection{Thermal equilibrium}
\label{Sec:thb}

\begin{figure*}
\begin{center}
\includegraphics[width=0.99\textwidth]{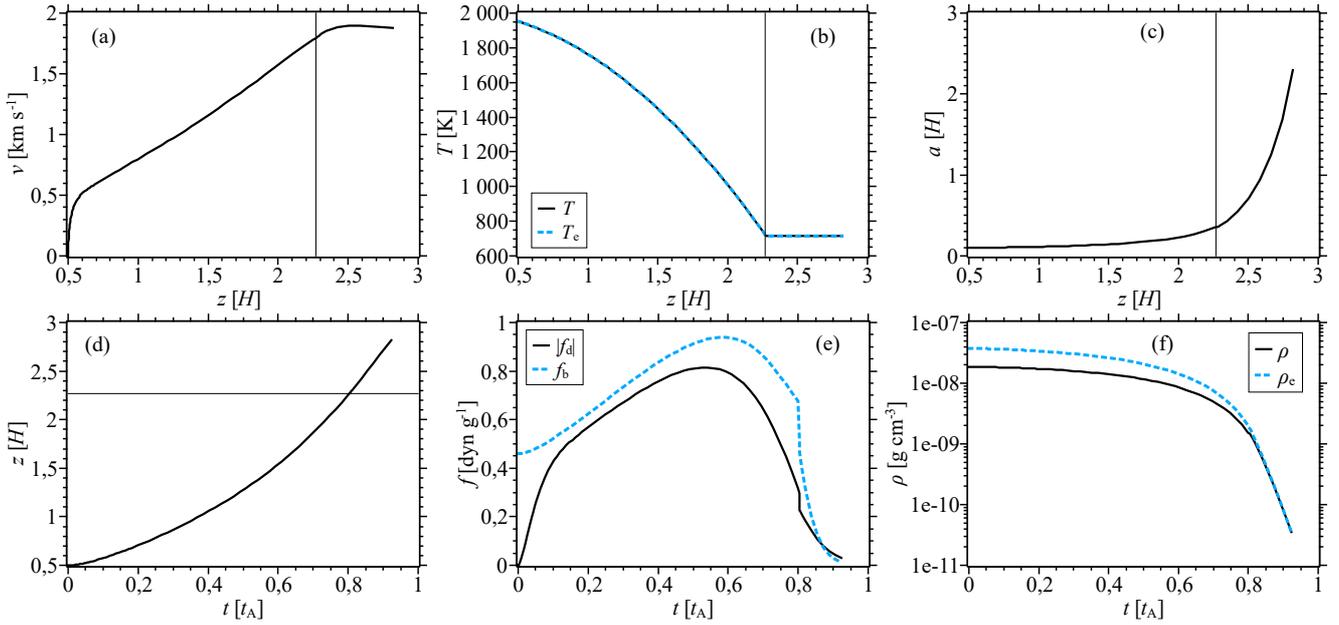}
\end{center}
\caption{Dynamics of the MFT in thermal equilibrium with ambient gas for $r=0.15$~au, $\beta_0=1$, $a_0=0.1H$, $z_0=0.5\,H$. Panel (a): vertical profile of velocity. Panel (b): vertical profile of internal and external temperatures (black solid and blue dashed lines, respectively). Panel~(c): vertical profile of MFT radius. Panel (d): dependence of MFT $z$-coordinate on time. Panel (e): dependence of  drag and buoyant forces on time (black solid and blue dashed lines, respectively). Absolute value of the drag force is depicted. Panel~(f): dependence of internal and external density on time (black solid and blue dashed lines, respectively). Vertical lines in panels (a-c) and horizontal line in panel (d) show the surface of the disc, $z_{{\rm s}}=2.27\,H$. The Alfv{\'e}n crossing time $t_{{\rm A}}=1.05\,P_{{\rm k}}$, where $P_{{\rm k}}$ is the Keplerian period equal to $0.06$~yr for the adopted parameters. (color figure online)}
\label{Fig:fiduc}
\end{figure*}

In this section, we discuss general features of the MFT dynamics in the case of thermal equilibrium. In Figures~\ref{Fig:fiduc}(a-c), we plot the dependences of velocity, temperature and radius of the MFT on its $z$-coordinate. Corresponding dependences of $z$-coordinate, drag and buoyant forces, internal end external densities on time are depicted in Figures~\ref{Fig:fiduc}(d-f). Absolute value of the drag force is plotted for convenience.

Figure~\ref{Fig:fiduc}(a) shows the MFT begins to rise with high acceleration. The velocity of the MFT grows very fast from 0 to $0.5\,\mathrm{km}\,\mathrm{s}^{-1}$,  because the buoyant force $f_{{\rm b}}$ is much stronger than the drag force $f_{{\rm d}}$ in the beginning of motion (see Figure~\ref{Fig:fiduc}(e)). After that, the buoyancy and drag forces become nearly equal to each other, $|f_{{\rm d}}|\lesssim f_{{\rm b}}$, and the velocity monotonically increases up to $\approx 1.8\,\mathrm{km}\,\mathrm{s}^{-1}$ at the surface of the disc $z_{{\rm s}}\approx 2.27\,H$. The MFT rises to the surface during time $\approx 0.8\,t_{{\rm A}} \approx 18$~d, as Figure~\ref{Fig:fiduc}(d) shows. Absolute values of the buoyant and drag forces become small, acceleration of the MFT vanishes and it acquires nearly steady velocity $v \approx 1.8\,\mathrm{km}\,\mathrm{s}^{-1}$ further.

The MFT is in thermal equilibrium with ambient gas, so that $T=T_{{\rm e}}$ during its motion, as Figure~\ref{Fig:fiduc}(b) demonstrates. Figure~\ref{Fig:fiduc}(c) shows  that the MFT expands in the course of the rise, i.e. its radius increases with $z$.  Correspondingly, density of the MFT decreases, as Figure~\ref{Fig:fiduc}(f) shows. Internal density always stays less than the external one in thermal equilibrium. Ultimately radius of the MFT exceeds the thickness of the disc, $a > z_{{\rm s}}$, at $z\approx 2.8\,H$. We do not simulate further motion of the MFT, as the slender tube approximation violates under such circumstances. The process of substantial expansion of MFT above the disc can be interpreted as a transformation of rising slender flux tubes into non-uniform expanding magnetized corona of the disc. Further we will call this process as a dispersal of the MFT. Effect of magnetic flux escape from the disc with subsequent formation of magnetized corona has been found by~\citet{miller00, machida00, johansen08, turner10, romanova11, takasao18} in the MHD simulations of the accretion discs. \citet{takasao18} found formation of the flux tubes from the regular magnetic field of the disc, that confirms our assumptions. 

\begin{figure*}
\begin{center}
\includegraphics[width=0.99\textwidth]{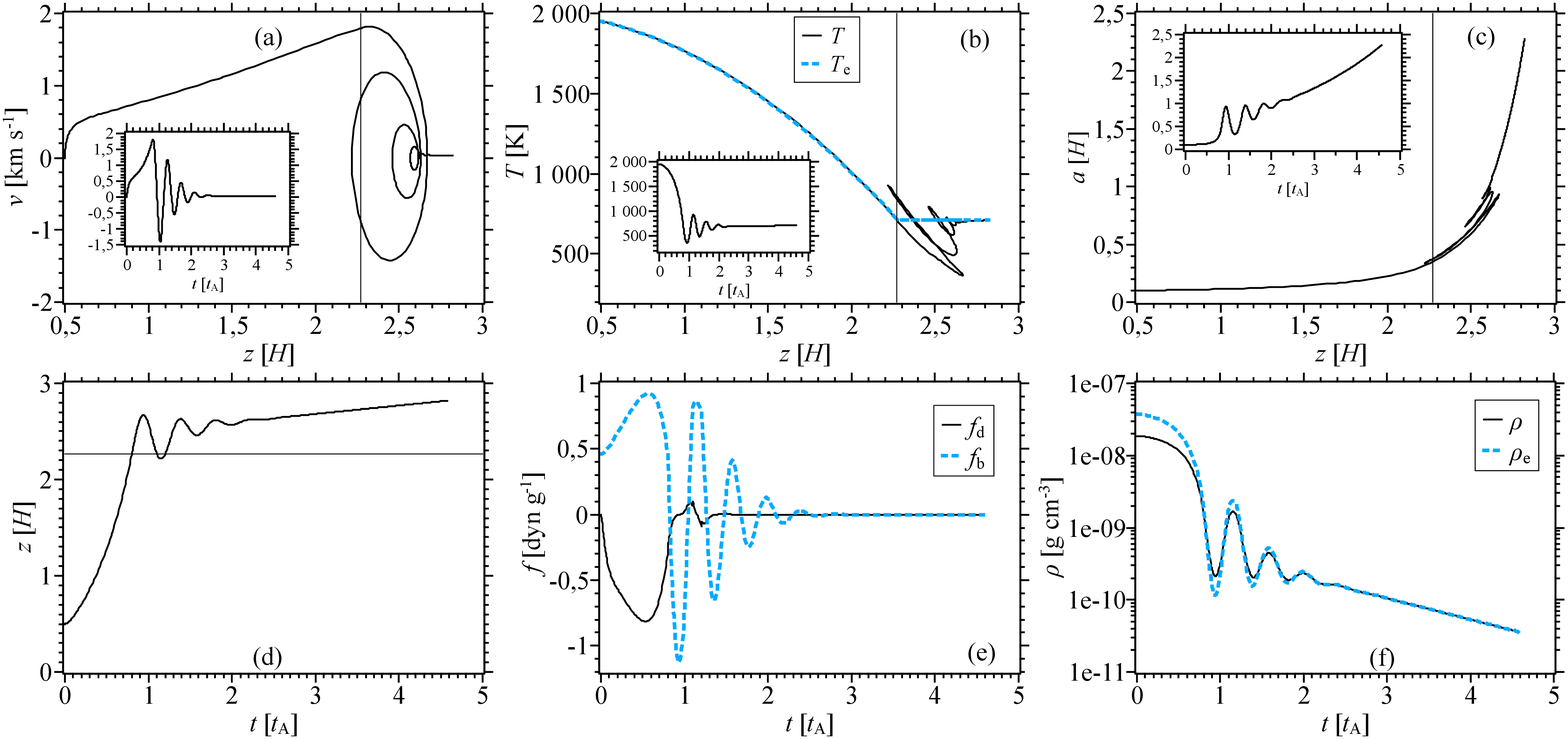}
\end{center}
\caption{Same as in Figure~\ref{Fig:fiduc}, but for the case when radiative heat exchange is taken into account. Insets in panels (a), (b) and (c) show dependence of MFT velocity, temperature and radius on time, respectively. (color figure online)}
\label{Fig:fiduc2}
\end{figure*}

\subsubsection{Role of heat exchange}
\label{Sec:he}
Consider dynamics of the MFT in the case, when radiative heat exchange is taken into account self-consistently (Figure~\ref{Fig:fiduc2}). Dynamics of the MFT inside the disc, $z<z_{{\rm s}}$ ($t<1\,t_{{\rm A}}$), is similar to the one discussed in Section~\ref{Sec:thb} (Figure~\ref{Fig:fiduc}). The MFT rises with increasing speed, expands and its density decreases.  It accelerates to $v\approx 1.8\,{\rm km}\,{\rm s}^{-1}$ near the surface of the disc. Temperature $T$ is nearly equal to $T_{{\rm e}}$, and its vertical profile matches the polytropic profile (\ref{Eq:Te}). Approximate equality of the temperatures inside and outside the MFT  reflects fast heat exchange at $z<z_{{\rm s}}$.

Figure~\ref{Fig:fiduc2}(a) shows that the MFT starts to decelerate after rising from the disc. The velocity goes to zero, when the MFT reaches point $z\approx 2.7\,H$ at $t\approx 1\,t_{{\rm A}}$. After that, the velocity becomes negative, and the MFT starts to move downwards, i.e. it `sinks'. When the MFT reaches point $z\approx 2.2\,H$, its velocity changes sign again, and the MFT starts to move upwards in the second time. Such an up and down motion is observed within the time period from  $t\approx 1\,t_{{\rm A}}$ to $t\approx 2\,t_{{\rm A}}$. Velocity periodically changes its sign, while its absolute value decreases at this part of the trajectory. Further, at $t>2\,t_{{\rm A}}$, the MFT rises from the disc and continues monotonic upward motion with small nearly steady velocity of about $0.04\,{\rm km}\,{\rm s}^{-1}$.

The oscillatory motion of the MFT near the surface of the disc in the time interval from $t\approx 1\,t_{{\rm A}}$ to $t\approx 2\,t_{{\rm A}}$ is explained by the following. The internal temperature of the MFT decreases with respect to constant temperature outside, when the MFT rises from the disc (see Figure~\ref{Fig:fiduc2}(b)). The  temperature takes minimum value $T_{{\rm min}}\approx 380$~K, when the MFT rise to the point $z\approx 2.7\,H$. The cooling of the MFT is caused by its practically adiabatic expansion. The radiative heat exchange appears to be inefficient to compensate adiabatic cooling at this part of the trajectory. Decrease of internal temperature with respect to the external one causes the MFT to expand more slowly than in the case of thermal equilibrium. The density of the MFT decreases with $z$ more slowly than the external density, and $\rho$ exceeds $\rho_{{\rm e}}$, i.e. the MFT loses buoyancy at the point $z\approx 2.4\,H$. The MFT moves by inertia upwards for some time, until it stops at $z\approx 2.7\,H$. Then the MFT starts to `sink' due to negative buoyancy. The MFT contracts a little, and its temperature and density grow in a process of downward motion (see Figures~\ref{Fig:fiduc2}(b, c, f)). The buoyancy restores ($\rho < \rho_{{\rm e}}$), when the internal and external temperatures become nearly equal to each other. This leads to deceleration of the MFT and to change of the velocity sign near the point $z=2.2\,H$. The radiative heat exchange leads to equalization of internal and external temperatures in a process of further periods of up and down motion with respect to the point of zero buoyancy ($\rho = \rho_{{\rm e}}$). Ultimately, temperatures $T$ and $T_{{\rm e}}$ become nearly equal to each other, and the MFT oscillations change onto monotonic upward motion with steady velocity above the disc surface at $t>2\,t_{{\rm A}}$. The oscillations decay due to equalization of the temperatures, and corresponding decrease of density difference and buoyant force. The MFT is optically thick during its motion. We call the oscillations discussed in this section as the thermal ones. 

\subsection{Dependence on parameters}
\label{Sec:Param}

We investigate the dynamics of MFT for various initial radii $a_0$, distance $r$ and plasma beta $\beta_0$ in this section. As it was mentioned in the introduction, MFT are likely form with $\beta_0\sim 1$. In this section we consider  MFT dynamics for plasma beta in range $[0.01,\,10]$ to study the dependence of MFT characteristics on the initial magnetic field strength. \citet{kh17raa} have shown that the characteristics of MFT near disc surface practically do not depend on the initial position $z_0$. In this section, we present calculations for $z_0=0.5H$. Radiative heat exchange is taken into account in all considered runs. Internal and external temperatures are equal to each other initially.

\begin{figure*}
\begin{center}
\includegraphics[width=0.99\textwidth]{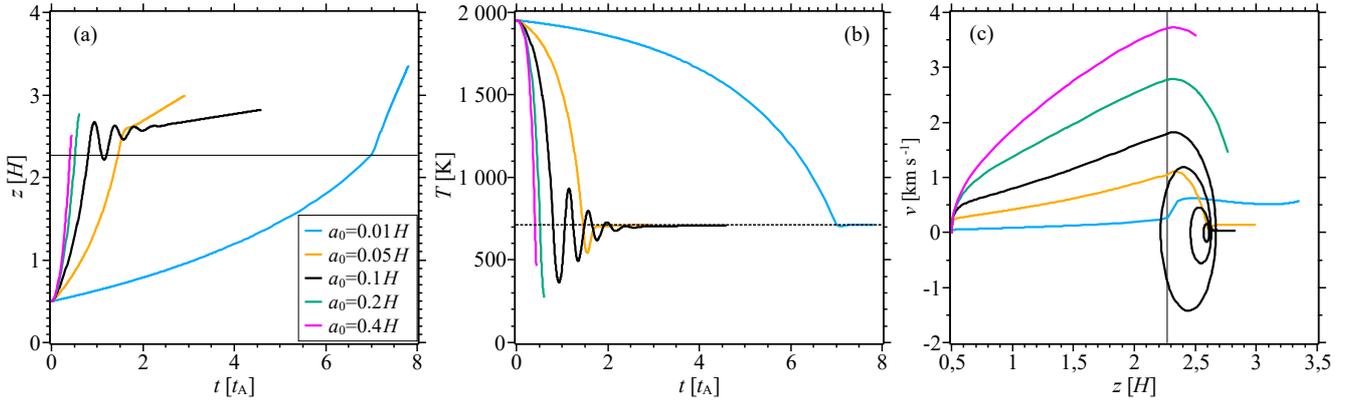}
\end{center}
\caption{Dynamics of the MFT in run with $r=0.15$~au, $\beta_0=1$, $z_0=0.5\,H$ and various initial radii $a_0$ (lines of different colors). Panel (a): dependence of MFT $z$-coordinate on time. Panel (b): dependence of MFT temperature on time. Panel (c): vertical profiles of MFT velocity. Horizontal line in panel (a) and vertical line in panel (c) depict the surface of the disc. Horizontal dashed line in panel (b) shows temperature of the gas above the disc, $T_{{\rm e}}=715$~K. The Alfv{\'e}n crossing time $t_{{\rm A}}=1.05\,P_{{\rm k}}$ and the Keplerian period $P_{{\rm k}}=0.06$~yr for the adopted parameters. (colour figure online)}
\label{Fig:param_a0}
\end{figure*}

In Figure~\ref{Fig:param_a0}, we plot the dependences of MFT $z$-coordinate and temperature on time, and velocity versus $z$-coordinate	, for $r=0.15$~au, $\beta_0=1$, $z_0=0.5\,H$ and various initial radii $a_0$. Black lines correspond to the fiducial case.

Figure~\ref{Fig:param_a0}(a) shows that rise time increases with decreasing radius of MFT. For example, the MFT with $a_0=0.01\,H$ (blue line) rises to the surface of the disc over time of $7\,t_{{\rm A}}$, while the MFT with $a_0=0.4\,H$ (magenta line) reaches disc surface within time of $0.4\,t_{{\rm A}}$ ($t_{{\rm A}}$ is  the Alfv{\'e}n crossing time). MFT with larger radius moves faster, because the buoyant force is proportional to MFT volume ($\propto a^3$), while the drag force is proportional to MFT surface area ($\propto a^2$) (see Figure~\ref{Fig:param_a0}(c)).

MFT with initial radius $a_0=0.1\,H$ in Figure~\ref{Fig:param_a0} exhibits thermal oscillations. Our simulations show that the MFT with initial radius $a_0>0.17\,H$ (see green and magenta lines in Figure~\ref{Fig:param_a0}) disperse fast above the disc before reaching the point of zero buoyancy. The MFT with initial radius $a_0<0.06\,H$ (see orange and blue lines in Figure~\ref{Fig:param_a0}) move slowly, so that the radiative heat exchange effectively equalizes internal and external temperatures (see Figure~\ref{Fig:param_a0}(b)), preventing the loss of buoyancy. 

\begin{figure}
\begin{center}
\includegraphics[width=0.49\textwidth]{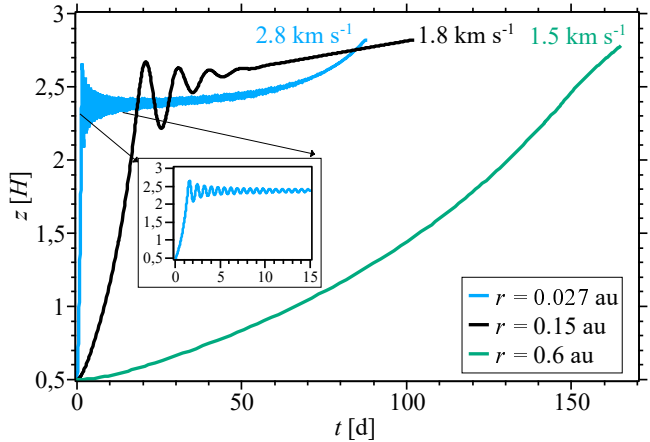}
\end{center}
\caption{Dependence of the MFT $z$-position on time in run with $a_0=0.1\,H$, $\beta_0=1$, $z_0=0.5\,H$ at various $r$-coordinates (lines of different colors). The inset shows zoomed-in region of the trajectory of the MFT for $r=0.027$~au. Numbers near the curves show maximum velocity of the MFT (blue for $r=0.027$~au, black for $r=0.15$~au, green for $r=0.6$~au). The Keplerian period equals $1.6$, $21$ and $170$~d at $r=0.027$, $0.15$ and $0.6$~au, respectively. (colour figure online)}
\label{Fig:param_r}
\end{figure}

In Figure~\ref{Fig:param_r}, we plot dependence of MFT $z$-coordinate on time for $a_0=0.1\,H$, $\beta_0=1$, $z_0=0.5\,H$ and various $r$-coordinates. Black lines correspond to the fiducial run discussed in Section~\ref{Sec:he}.

Figure~\ref{Fig:param_r} shows that time of MFT rise to the surface of the disc increases with $r$-distance. The MFT float up to the surface over time of $2$~d at $r=0.027$~au, and $140$~d at $r=0.6$~au, and velocity of the MFT decreases from $2.8\,{\rm km}\,{\rm s}^{-1}$ to $1.5\,{\rm km}\,{\rm s}^{-1}$. 

MFT experience the thermal oscillations only in the region $r\leq 0.2$~au (blue and black lines in Figure~\ref{Fig:param_r}). Oscillation period increases with $r$-distance. It is approximately equal to $1$~d at the distance $r=0.027$~au (blue line), and $10$~d at the distance $r=0.15$~au (black line). The simulations of MFT dynamics for various initial plasma beta show that only the MFT with $\beta_0=1$ experience thermal oscillations in the region $r\leq 0.2$~au.

In the region $r>0.2$~au, the MFT rise so slowly that the radiative heat exchange is able to equalize internal and external temperatures, preventing the loss of buoyancy.

In Table \ref{Tab:B}, we present initial parameters and some results of the simulations of the MFT dynamics at  the distances $r=0.027$ and $r=0.6$~au.  Initial coordinate $z_0$ is $0.5\,H$ in all presented runs. Column~1 shows number of run. Column~2 contains the values of $r$-distance. Initial plasma beta $\beta_0$ and radius of the MFT $a_0$ are listed in columns 3 and 4, respectively. The MFT mass, $M$, and magnetic flux, $\Phi$, are given in columns 5 and 6, respectively. We list velocity, $v_{{\rm surf}}$, radius, $a_{{\rm surf}}$, and magnetic energy, $E_{{\rm m}}$, of the MFT in the moment, when it crosses the disc surface, in columns 7, 8 and 9. Time of rise to the surface of the disc, $t_{{\rm surf}}$, is given in column 10. Column 11 gives time of the toroidal magnetic field generation (see discussion in Sections~\ref{Sec:outflows} below). Column 12 contains the rate of mass loss due to buoyancy (see discussion in Section~\ref{Sec:outflows} below). In column 13, we list ratios of the minimal MFT temperature during its rise with respect to corresponding external temperature. Typically, the minimal temperature is achieved at $z\approx 2.3-3\,H$.

\begin{table*}
\centering	
\caption{Parameters of runs \label{Tab:B}}
\begin{tabular}{@{}cccc|ccccccccc@{}}
\hline
run & $r$ [au] & $\beta_0$ & $a_0$ [$H$] & $M$ [$M_{\sun}$] & $\Phi$ [Mx] & $v_{{\rm surf}}$ [${\rm km}\,{\rm s}^{-1}$] & $a_{{\rm surf}}$ [$H$] &  $E_{{\rm m}}$, [erg] &  $t_{{\rm surf}}$ [yr] & $t_{{\rm gen}}$ [yr] & $\dot{M}_{{\rm b}},\,[M_{\sun}\,{\rm yr}^{-1}]$  & $T_{{\rm min}}/T_{\rm e}$\\ 
(1) & (2) & (3) & (4) & (5) & (6) & (7) & (8) & (9) & (10) & (11) & (12) & (13)\\
\hline
1   & 0.6 & 0.01		& 0.01	&  $2.8\times 10^{-11}$& $2.9\times 10^{21}$ &  0.6	& 0.034 & $2.2\times 10^{34}$&2.0	& 10.6 & $2.6\times 10^{-12}$ & 1.0\\ 
2   & 0.6 & 0.1	    & 0.01	&  $2.5\times 10^{-10}$& $2.7\times 10^{21}$	& 0.5	& 0.035	&  $1.7\times 10^{34}$& 2.2	& 10.2 & $2.5\times 10^{-11}$ & 1.0\\ 
3   & 0.6 & 1			& 0.01	&  $1.4\times 10^{-9}$	& $2.0\times 10^{21}$	& 0.2	& 0.046	& $5.5\times 10^{33}$& 3.5	& 7.6 & $1.8\times 10^{-10}$  & 1.0\\ 
4   & 0.6 & 10		& 0.01	&  $2.5\times 10^{-9}$	& $8.6\times 10^{20}$	& 0.05	& 0.059	 & $4.8\times 10^{34}$& 12.0	& 3.2 & $1.8\times 10^{-10}$  & 1.0 \\ 
5   & 0.6 & 0.01		& 0.1	&  $2.8\times 10^{-9}$	& $2.9\times 10^{23}$	& 5.2	& 0.336	& $2.1\times 10^{36}$& 0.2	& 10.6 & $2.6\times 10^{-10}$  & 1.0\\ 
6   & 0.6 & 0.1		& 0.1	&  $2.5\times 10^{-8}$	& $2.7\times 10^{23}$  & 3.6	& 0.355	 &$1.7\times 10^{36}$& 0.2 & 10.2 & $2.5\times 10^{-9}$  & 0.95\\ 
7   & 0.6 & 1			& 0.1	&  $1.4\times 10^{-7}$	& $2.0\times 10^{23}$	& 1.5	& 0.463	 & $5.6\times 10^{35}$& 0.4	& 7.6 & $1.8\times 10^{-8}$  & 0.78 \\ 
8   & 0.6 & 10		& 0.1	&  $2.4\times 10^{-7}$	& $8.6\times 10^{22}$	& 0.3	& 0.578	 & $6.6\times 10^{34}$& 1.4	& 3.2 & $7.9\times 10^{-8}$ & 0.91 \\ 

9   & 0.6 & 0.01		& 0.2	&  $1.1\times 10^{-8}$	& $1.1\times 10^{24}$	& 9.0	& 0.673	& $8.4\times 10^{36}$& 0.1	& 10.6 & $1.0\times 10^{-9}$  & 1.0\\ 
10   & 0.6 & 0.1		& 0.2	&  $1.0\times 10^{-7}$	& $1.1\times 10^{24}$	& 5.5	& 0.709	 & $7.0\times 10^{36}$& 0.1	& 10.2 & $9.8\times 10^{-9}$ & 0.83\\ 
11   & 0.6 & 1			& 0.2	&  $5.6\times 10^{-7}$	& $8.1\times 10^{23}$	& 2.2	& 0.925	 & $2.2\times 10^{36}$& 0.3	& 7.6 & $7.4\times 10^{-8}$  & 0.58 \\ 
\hline 
12   & 0.027 & 0.01	& 0.01	& $6.3\times 10^{-13}$	& $7.0\times 10^{19}$  & 0.9	& 0.028	 &$2.8\times 10^{33}$& 0.02	& 0.69 & $9.1\times 10^{-13}$ &1.0 \\ 
13   & 0.027 & 0.1		& 0.01	& $5.8\times 10^{-12}$	& $6.8\times 10^{19}$	& 0.8	& 0.029	 & $2.4\times 10^{34}$& 0.02	& 0.66 & $8.8\times 10^{-12}$ &1.0\\ 
14   & 0.027 & 1			& 0.01	& $3.2\times 10^{-11}$	& $5.0\times 10^{19}$	& 0.4	& 0.036	 & $8.7\times 10^{32}$& 0.03	& 0.49 & $6.5\times 10^{-11}$  &1.0 \\ 
15   & 0.027 & 10			& 0.01	& $3.2\times 10^{-11}$	& $2.3\times 10^{19}$	& 0.07	& 0.047	 & $1.0\times 10^{32}$& 0.10	& 0.20 & $2.9\times 10^{-10}$  & 0.98\\
16   & 0.027 & 0.01	& 0.1	& $6.3\times 10^{-11}$	& $7.0\times 10^{21}$	& 8.6	& 0.276	 &  $2.8\times 10^{35}$& 0.002	& 0.69 & $9.1\times 10^{-11}$ & 0.98\\ 
17 & 0.027 & 0.1		& 0.1	& $5.8\times 10^{-10}$	& $6.8\times 10^{21}$	& 6.5	& 0.290	 &  $2.3\times 10^{35}$& 0.002	& 0.66 & $8.8\times 10^{-10}$ & 0.34\\ 
18   & 0.027 & 1			& 0.1	& $3.2\times 10^{-9}$	& $5.0\times 10^{21}$	& 2.8	& 0.356	 & $8.4\times 10^{34}$& 0.004	& 0.49  & $6.5\times 10^{-9}$  & 0.48\\ 
19  & 0.027 & 10			& 0.1	& $5.8\times 10^{-9}$	& $2.3\times 10^{21}$	& 0.5	& 0.431	 & $1.0\times 10^{34}$& 0.13	& 0.2  & $2.9\times 10^{-8}$  & 0.89\\ 
20   & 0.027 & 0.01	& 0.2	& $2.5\times 10^{-10}$	&  $2.8\times 10^{22}$	& 15.6	& 0.554	 & $1.1\times 10^{36}$& 0.001	& 0.69 & $3.6\times 10^{-10}$  & 0.58 \\ 
21   & 0.027 & 0.1		& 0.2	& $2.3\times 10^{-9}$	& $2.7\times 10^{22}$	& 10.3	& 0.576	 & $9.4\times 10^{35}$& 0.0011	& 0.66 & $3.5\times 10^{-9}$  & 0.35 \\ 
22   & 0.027 & 1			& 0.2	& $1.3\times 10^{-8}$	& $2.0\times 10^{22}$	& 4.3	& 0.708	& $3.4\times 10^{35}$& 0.0024	& 0.49 & $2.7\times 10^{-8}$  & 0.39\\ 

\hline 
\end{tabular} 
\end{table*}

Table \ref{Tab:B} shows that masses of MFT range from $6.3\times 10^{-13}\,M_{\sun}$ (run 12) to $1.0\times 10^{-7}\,M_{\sun}$ (run 10). The MFT mass increases with its radius. The density and, therefore, MFT mass increase with $\beta_0$ (it follows from Equation (\ref{Eq:drho})). The MFT transfer magnetic fluxes in range from $2.3\times 10^{19}$~Mx (run 15) to $1.1\times 10^{24}$~Mx (run 9). Velocity of the MFT with $\beta_0=1$ has the values in interval $0.2-4.3\,{\rm km}\,{\rm s}^{-1}$ that is comparable with local sound speed of $1-2.5\,{\rm km}\,{\rm s}^{-1}$. The MFT with smaller $\beta_0$ accelerate to higher velocities and can have supersonic speed. Maximum velocity $\sim 15.6\,{\rm km}\,{\rm s}^{-1}$ is achieved in run 20. In this case, bow shocks will probably form during motion of the MFT, which can change drag law and lead to additional heating. In the following sections we will present and discuss the simulations, in which the velocity of the MFT does not exceed significantly the sound speed.

When the MFT crosses the disc surface, its radius is several times smaller than $z_{{\rm s}}$. For example, $a(z=z_{{\rm s}})=0.035\,H$ for run 1. The MFT continue to expand, when they move above the disc (see Figure~\ref{Fig:fiduc}(c)). The MFT radius becomes larger than the disc height at $z\sim 2.7-3.5\,H$ in performed runs. For instance,  the MFT radius exceeds the disc height at $z=2.7\,H$ in run~8, and at $z=3.5\,H$ in run 9. As it was stated in Section~\ref{Sec:Fiduc}, we interpret the dispersal of the rising MFT as a formation of expanding magnetized corona above the surface of the disc. 

Thinnest MFT with $a_0\leq 0.01\,H$ stay in thermal equilibrium with external gas, $T\approx T_{{\rm e}}$, as column 13 of Table~\ref{Tab:B} shows. Thicker MFT cool down in comparison to the external gas during their motion. For example, the MFT with $\beta_0=1$, $a_0=0.2$ and $r=0.6$~au cools down to minimal temperature $\approx 0.58 \,T_{{\rm e}}$ (run 11). Among the runs presented in Table \ref{Tab:B}, prominent thermal oscillations are found only in runs 18 and 19, in agreement with the discussion of the Figure~\ref{Fig:param_r}. Oscillation behaviour is also observed in run 8 with $\beta_0=10$, $a_0=0.2$, $r=0.6$, but the oscillations rapidly decay within three periods in this case. In the other runs, the MFT disperse fast above the disc.

\subsection{Mass and magnetic flux loss due to buoyancy of flux tubes}
\label{Sec:outflows}

As it has been shown in sections~\ref{Sec:Fiduc} and \ref{Sec:Param}, the MFT rise from the disc to its atmosphere carrying away mass and magnetic flux.  The rising MFT can be the seed for the formation of jets and outflows from accretion discs. Similar idea was considered by \cite{chakra94} and \cite{deb17} in application to the accretion discs around black holes.  The rate of vertical mass transport via buoyancy can be estimated as a mass of the flux tubes rising from the disc per unit of time. We consider times of the MFT formation and rise as characteristic time scales. The characteristic time of MFT formation is comparable to the time of the toroidal magnetic field generation. To estimate efficiency of the mass transport form disk interior to its atmosphere  via buoyancy, we compare the characteristic times in this section.

The time of the azimuthal magnetic field generation can be estimated from the $\varphi$-component of induction equation in the approximations of the accretion disc model,
\begin{equation}
	\frac{\upartial B_{\varphi}}{\upartial t} = B_z\frac{\upartial v_{\varphi}}{\upartial z}.\label{Eq:IndPhi}
\end{equation}
This equation shows that $B_{\varphi}$ is generated from $B_z$ by the differential rotation of the disc. The velocity $v_{\varphi}$ is determined from the balance between the gravity and centrifugal force in the $r$-direction. In our case,
\begin{equation}
	v_{\varphi} = \sqrt{\frac{GM_{\star}}{r}}\left(1 + \frac{z^2}{r^2}\right)^{-3/4} = r\Omega_{{\rm k}}\left(1 + \frac{z^2}{r^2}\right)^{-3/4},
\end{equation}
so Equation (\ref{Eq:IndPhi}) in the case $z^2/r^2 \ll 1$ turns to
\begin{equation}
	\frac{\upartial B_{\varphi}}{\upartial t} \simeq -\frac{3}{2}\frac{z}{r}B_z\Omega_{{\rm k}}.\label{Eq:IndPhi2}
\end{equation}
Therefore, the time of the azimuthal magnetic field generation up to a given value $B_{\varphi}$
\begin{equation}
	t_{{\rm gen}} = \frac{2}{3}\frac{B_{\varphi}}{B_z}\Omega_{{\rm k}}^{-1}\left(\frac{z}{r}\right)^{-1}\simeq 2.12P_{{\rm k}}\left(\frac{z/r}{0.05}\right)^{-1}\frac{B_{\varphi}}{B_z},\label{Eq:t_phi}
\end{equation}
where $P_{{\rm k}}$ is the Keplerian period. Typical time scale of $B_{\varphi}$ amplification up to the value $B_{\varphi} = B_z$  is nearly two Keplerian periods for $z=0.05\,r$. Keplerian period increases with distance as $P_{{\rm k}}\propto r^{3/2}$, therefore fastest generation of $B_{\varphi}$ takes place in the innermost region of the disc.

\begin{figure}
\begin{center}
\includegraphics[width=0.49\textwidth]{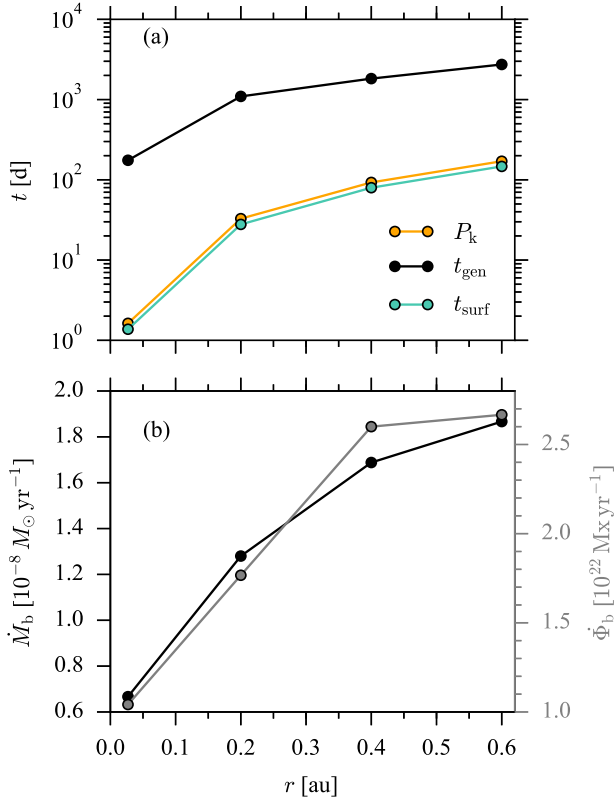}
\end{center}
\caption{Panel (a): dependences of Keplerian period $P_{{\rm k}}$ (grey line), MFT generaton time $t_{{\rm gen}}$ (orange line) and MFT rise time  $t_{{\rm surf}}$ (green line) on the $r$-coordinate. Panel (b): dependences of  the rates of vertical mass transport (black line, left $y$-axis) and magnetic flux transport (grey line, right $y$-axis) due to buoyancy on the $r$-coordinate. Initial parameters of the MFT: $a_0=0.1\,H$, $\beta_0=1$. (colour figure online)}
\label{Fig:Mdot_b}
\end{figure}

MFT rise and generation times for considered parameters are listed in columns 10 and 11 of Table~\ref{Tab:B}, respectively. We plot dependence of $P_{{\rm k}}$, $t_{{\rm gen}}$ and  $t_{{\rm surf}}$ on the $r$-coordinate in Figure~\ref{Fig:Mdot_b}(a). Table~\ref{Tab:B} and Figure~\ref{Fig:Mdot_b}(a) show that characteristic times increase with distance. The rise time $t_{{\rm surf}}$ increases from 0.5~d at $0.027$~au to $200$~d at $r=0.6$~au. It is nearly equal to Keplerian period and order of magnitude less than generation time $t_{{\rm gen}}$ (\ref{Eq:t_phi}). This means that the dynamics of the MFT will occur in two stages. At the first stage having duration $t_{{\rm gen}}$, toroidal magnetic field is generated and MFT form due to magnetic buoyancy instability. At the second stage, the MFT rise from the disc over the time $t_{{\rm surf}}$ and carry away some mass and magnetic flux to disc atmosphere. After that, this two-stage process repeats. Since $t_{{\rm surf}}\ll t_{{\rm gen}}$, the process of mass and magnetic flux transport from disc to its atmosphere is periodic with typical period $t_{{\rm gen}}$. Table~\ref{Tab:B} and Figure~\ref{Fig:Mdot_b}(a) show that $t_{{\rm gen}}\approx (0.5-0.7)$~yr at $r=0.027$~au and $t_{{\rm gen}}\approx (8-10)$~yr at $r=0.6$~au. 

The MFT can form inside the region $r=0.027\div 0.6$~au. Results presented in Figure~\ref{Fig:Mdot_b}(a) indicate that MFT will form inside the disc and rise from it first at small $r$  and then at farther distances from the star.

The rate of vertical mass transport via the buoyancy, $\dot{M}_{{\rm b}}$,  can be estimated by division of the MFT mass by the characteristic time of the magnetic field amplification $t_{{\rm gen}}$. Values of $\dot{M}_{{\rm b}}$ lie in range $\sim 10^{-12}-10^{-7}\,M_{\sun}\,{\rm yr}^{-1}$ (column 12 of Table~\ref{Tab:B}). This value increases with plasma beta and radius of MFT. For example, maximum rate $\dot{M}_{{\rm b}}=7.9\times 10^{-8}\,M_{\sun}\,{\rm yr}^{-1}$ is found in run 8 with $a_0=0.1\,H$ and $\beta_0=10$.  
Rate of magnetic flux transport via buoyancy can be estimated in similar way, $\dot{\Phi}_{{\rm b}}=\Phi / t_{{\rm surf}}$. 

In Figure~\ref{Fig:Mdot_b}(b), we plot the dependences of $\dot{M}_{{\rm b}}$ and  $\dot{\Phi}_{{\rm b}}$ on the $r$-coordinate for the MFT with initial parameters $a_0=0.1\,H$, $\beta_0=1$, $z_0=0.5\,H$.  Figure~\ref{Fig:Mdot_b}(b) shows that $\dot{M}_{{\rm b}}$ increases with distance, from $\approx 6\times 10^{-9}\,M_{\odot}\,{\rm yr}^{-1}$ at $r=0.027$~au to $\approx 2\times 10^{-8}\,M_{\odot}\,{\rm yr}^{-1}$ at $r=0.6$~au. Therefore, the periodic process of mass and magnetic flux transport from the disc to its atmosphere caused by magnetic buoyancy is the most efficient near the outer edge of thermal ionization zone, where the MFT with large radius form. The rate of vertical mass transport via the buoyancy at $r=0.6$~au is five times smaller than the mass accretion rate in the disc $\dot{M}$, i.e. $20\%$ of the accreted mass can be transported from the disc to its atmosphere via buoyancy.

It should be noted that `dead' zone is situated in the region between $r=0.6$~au and $r=33$~au for the considered parameters of the disc. The rate $\dot{M}_{{\rm b}}$ will decrease rapidly with $r$ beyond $r=0.6$~au, since amplification of the toroidal magnetic field is hindered by Ohmic diffusion and MFT cannot form inside the `dead' zone. MFT can form only in the surface layer of the disc above the `dead' zone. Typical surface density of this layer is $5-10\,{\rm g\,cm}^{-2}$ (e.g., \citet{fmfadys}), and the MFT forming inside this layer have small radius of $\sim 0.01\,H$ and carry away tiny mass.

Figure~\ref{Fig:Mdot_b} shows that rate of magnetic flux transport via buoyancy also increases with $r$, from $\dot{\Phi}_{{\rm b}}\approx 1\times 10^{22}\,{\rm Mx\,yr}^{-1}$ at $r=0.027$~au to $\dot{\Phi}_{{\rm b}}\approx 2.7\times 10^{22}\,{\rm Mx\,yr}^{-1}$ at $r=0.6$~au. Minimum and maximum magnetic fluxes are $2.3\times 10^{19}$ and $1.1\times 10^{24}$~Mx (runs 15 and 9), and magnetic energies are $3.0\times 10^{33}$ and $3\times 10^{37}\,{\rm erg}$ (see columns 6 and 9 of Table~\ref{Tab:B}). Total magnetic flux of the disc equals $5\times 10^{29}$~Mx for considered parameters. Therefore, nearly $20\%$ of the disc magnetic flux can be lost via the magnetic buoyancy over $\sim 1$~Myr.

\subsection{Effect of external magnetic field}
\label{Sec:oscill}

In sections~\ref{Sec:Fiduc}-\ref{Sec:outflows}, we investigated the dynamics of the MFT in the disc without external magnetic field. The magnetic field of the disc can influence the MFT dynamics mainly through the magnetic pressure. 
In this section, we discuss the MFT dynamics in the case, when the magnetic pressure outside the MFT is taken into account in pressure equilibrium Equation~(\ref{Eq:pbal2}), i.e.
\begin{equation}
	P + \frac{B^2}{8\upi} = P_{{\rm e}} + \frac{B_{{\rm e}}^2}{8\upi}.\label{Eq:Pe_mag}
\end{equation}

\begin{figure*}
\begin{center}
\includegraphics[width=0.95\textwidth]{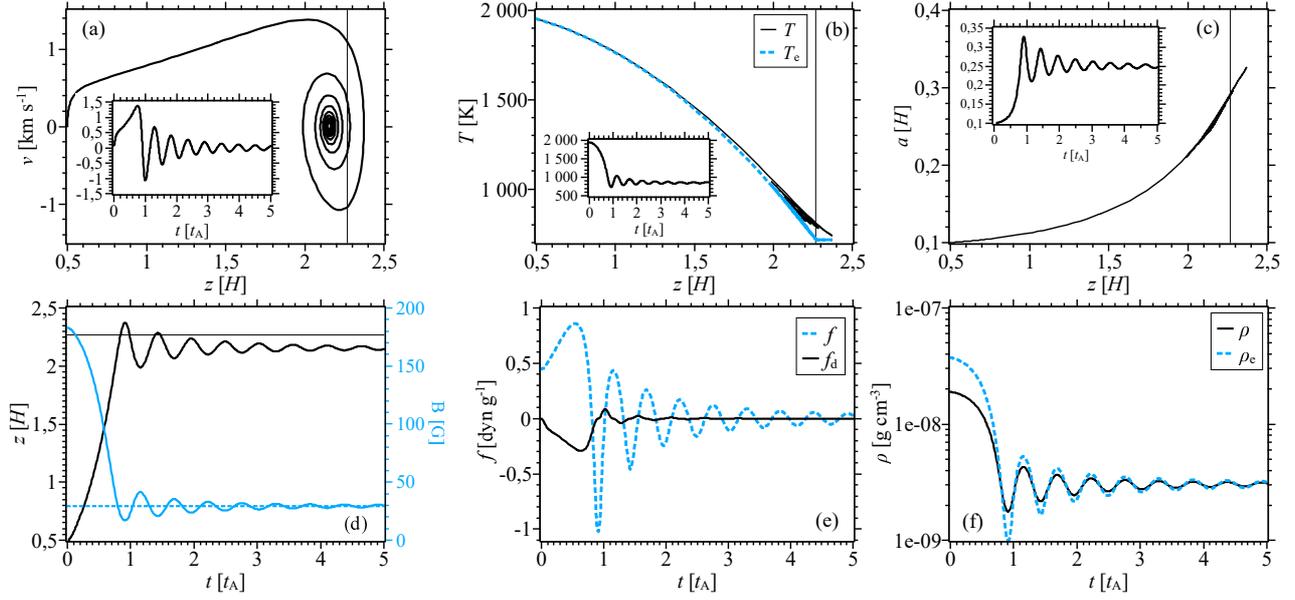}
\end{center}
\caption{Same as in Figure~\ref{Fig:fiduc}, but for the case, when effect of the external magnetic field is taken into account. Magnetic field strength of the MFT is depicted in panel (d) with the solid blue line (right $y$-axis). The horizontal dashed blue line in panel (d) shows the strength of external magnetic field. (colour figure online)}
\label{Fig:mf_fiduc}
\end{figure*}

First, we discuss the dynamics of the MFT for the fiducial parameters $r=0.15$~au, $\beta_0=1$, $a_0=0.1H$, $z_0=0.5\,H$ (see Section~\ref{Sec:thb}). 
In Figure~\ref{Fig:mf_fiduc}, we plot the vertical profiles of velocity, temperature and radius of the MFT (panels (a), (b) and (c), respectively), as well as dependence of its $z$-coordinate, drag and  buoyant forces, internal and external densities on time (panels (d), (e) and (f), respectively). Insets in panels (a-c) show dependences of $v$, $T$ and $a$ on time. Simulation is performed for the case of thermal equilibrium to eliminate effects of thermal oscillations.

Figure~\ref{Fig:mf_fiduc}(a) shows that the MFT floats with increasing velocity at the initial part of rise, $z<2\,H$ ($t<0.75\,t_{{\rm A}}$). Velocity of the MFT reaches the value of $1.35\,{\rm km}\,{\rm s}^{-1}$ at $z=2\,H$. After that the MFT decelerates, and its velocity goes to zero at $z=2.35\,H$. Then the velocity becomes negative, and the MFT starts to move downwards. Its velocity changes sign again at $z=2\,H$, and the MFT starts to move upwards. In the following, the MFT experiences oscillatory motion up and down with respect to the point $z\approx 2.15\,H$. Maximum value of the velocity decreases during these oscillations, i.e. the oscillations decay. Figures~\ref{Fig:mf_fiduc}(b, c) show that the MFT pulsates in the process of the oscillations, i.e. its temperature and radius periodically change with respect to the values of $860$~K and $0.25\,H$, respectively. Period of the oscillations, $P_{{\rm osc}}$, equals $0.4\,t_{{\rm A}}$.

The oscillations of the MFT are explained as follows. Figures~\ref{Fig:mf_fiduc}(e, f) demonstrate that at the initial part of the trajectory,  $t<0.75\,t_{{\rm A}}$,  the MFT density is less than the external density, and the buoyant force is positive. Figures~\ref{Fig:mf_fiduc}(c, d) show that the MFT expands and its magnetic field weakens. Strength of the external magnetic field, $B_{{\rm e}}$, is assumed to be constant in our simulations. Internal magnetic field strength $B$ becomes less than $B_{{\rm e}}$ at the point $z=2.18\,H$ ($t=0.8\,t_{{\rm A}}$). Increasing magnetic pressure outside the MFT causes it to expand slower than in the case of rise without external magnetic field (see Figure~\ref{Fig:fiduc}(f)). Density of the MFT is greater than the external density, and the buoyant force is negative, since $B$ is less than $B_{{\rm e}}$ in this point. The MFT rise to the point $z\approx 2.35\,H$ by inertia,  and then start to move downwards. The MFT contracts during the downward motion, i.e. its radius decreases, while the temperature, density and magnetic field strength increase. The  buoyancy become positive again, and the MFT starts to move upwards after $B$ exceeds $B_{{\rm e}}$. In the following, the MFT oscillates near the point of zero buoyancy, that is determined by equality $B=B_{{\rm e}}$ in this case. We call these oscillations as magnetic ones.

Drag force reduces the kinetic energy of the MFT, and leads to the decay of oscillations. Magnitude of the velocity oscillations reduces by factor of ten over five periods of oscillations. The MFT radius approaches the value of $0.25\,H$ in this case.

The magnetic oscillation has one significant difference form the thermal ones discussed in Section~\ref{Sec:he}. The thermal oscillations are caused by departure from thermal equilibrium. Therefore, they cease, and the MFT start to move upwards monotonically after the temperatures inside and outside the MFT become equal to each other. The magnetic oscillations are caused by the action of the external magnetic pressure. In the considered case, constant $B_{{\rm e}}$ prevents further rise of the MFT. Dynamics of the MFT is characterized by decaying oscillations near the point of zero buoyancy. In the case, when external magnetic field $B_{{\rm e}}$ decreases with height, the MFT would float farther.

We performed the simulation for the same parameters as discussed above, but taking into account the radiative heat exchange. The simulations show that the picture of the MFT dynamics is practically the same, as in Figure~\ref{Fig:mf_fiduc}, i.e. departure from thermal equilibrium does not influence the magnetic oscillations.

\begin{figure*}
\begin{center}
\includegraphics[width=0.95\textwidth]{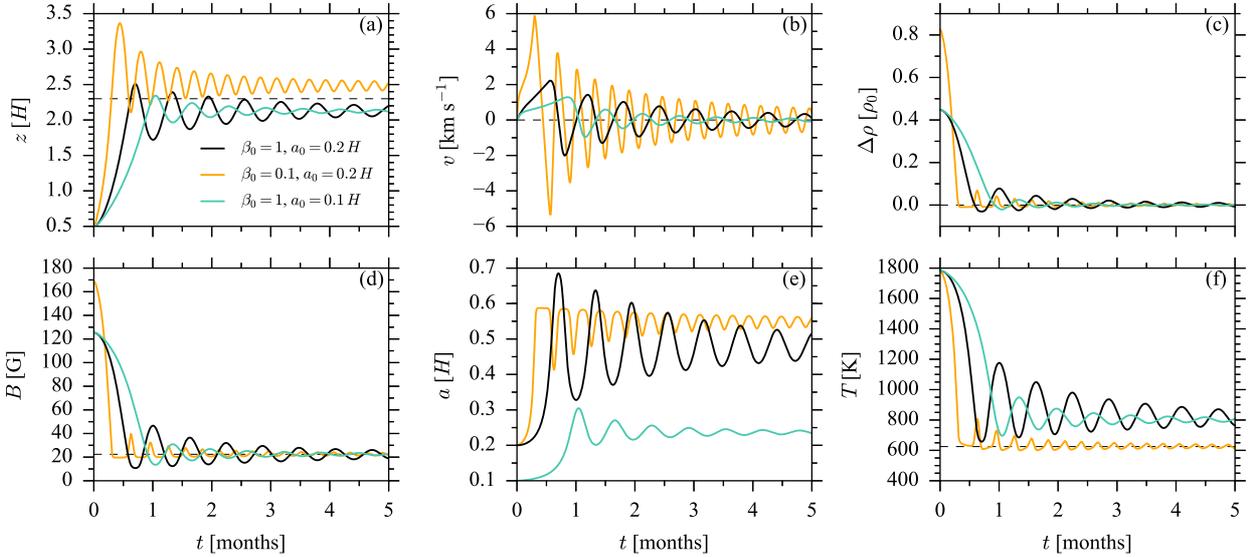}
\end{center}
\caption{The dependences of $z$-coordinate  (panel (a), horizontal dashed line is the surface of the disc), velocity (panel (b)), density difference (panel (c)), magnetic field strength (panel (d), horizontal dashed line is the external magnetic field $B_{{\rm e}}$), radii (panel (e)) and temperature $T$ (panel (f), horizontal dashed line is the external temperature above the disc $T_{{\rm a}}$) on time for the case when external magnetic pressure is taken into account. Black line: $\beta_0=1$ and $a_0=0.2\,H$, grey line: $\beta_0=0.1$ and $a_0=0.2\,H$, green line: $\beta_0=1$ and $a_0=0.2\,H$. Distance to the star is $r=0.2$~au. The Alfv{\'e}n crossing time $t_{{\rm A}}=0.94\,P_{{\rm k}}$ and the Keplerian period $P_{{\rm k}}=2.7$~months for the adopted parameters. (colour figure online)}
\label{Fig:param_b}
\end{figure*}

Figure~\ref{Fig:param_b} demonstrates dynamics of the MFT taking into account external magnetic field for various initial radii and plasma betas at $r=0.2$~au. 

Figure~\ref{Fig:param_b} shows that the magnetic oscillations are observed for all considered parameters. The MFT with stronger magnetic field (small $\beta_0$) experience oscillations at higher altitude. For example, the MFT with $\beta_0=0.1$ and $a_0=0.2\,H$ (orange lines) oscillate above the surface of the disc, near the point $z=2.5\,H$. Oscillation period in this case is less than in the case $\beta_0=1$ (black lines), while magnitude of velocity oscillations is higher, and magnitude of radius variations is smaller.

The MFT with a smaller initial radius move slower, but oscillate at about the same height as the tubes of a larger radius at the same initial field strength (compare black and green lines in Figure~\ref{Fig:param_b}).

We carried out the simulations of MFT dynamics taking into account external magnetic pressure for a full set of parameters $a_0$ and $\beta_0$ considered in Table~\ref{Tab:B} and for various $r$-coordinates in range $[0.027,\,0.6]$~au. The simulations show that the altitude, at which MFT oscillate, increases with initial magnetic field strength of the MFT. The oscillations of the MFT with $\beta_0=1$ take place under the surface of the disc, while the MFT with $\beta_0<1$ oscillate above the surface. Typical radii of the MFT are of order of $0.5\,H$, so that the oscillations of the MFT will lead to periodical changes of the disc structure near its surface, in the region from $z\approx 1.5\,H$ to $z\approx 3\,H$.

\begin{figure}
\begin{center}
\includegraphics[width=0.49\textwidth]{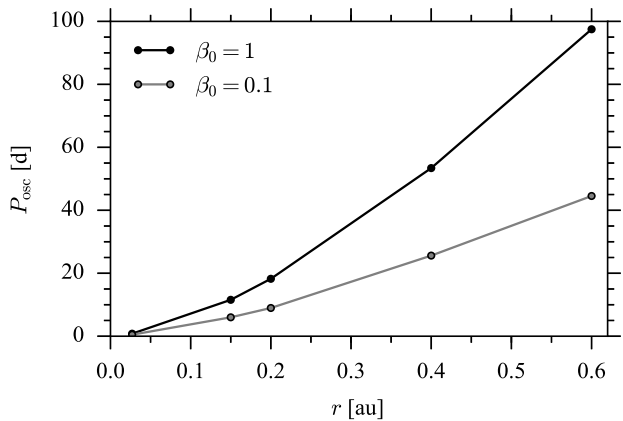}
\end{center}
\caption{Dependences of the period of the MFT' oscillations on the $r$-coordinate. Unit of time is day. Initial parameters of the MFT: $a_0=0.1\,H$, $z_0=0.5\,H$. Black line: $\beta_0=1$, grey line: $\beta_0=0.1$.}
\label{Fig:Period}
\end{figure}

In Figure~\ref{Fig:Period}, we plot the periods of MFT oscillations $P_{{\rm o}}$ for various $r$-coordinates. The results are obtained for the MFT with initial radius $a_0=0.1\,H$, coordinate $z_0=0.5\,H$, plasma beta $\beta_0=1$ (black line) and $\beta_0=0.1$ (grey line). Figure~\ref{Fig:Period} shows that oscillation period increases with the radial distance from $1$~d $\approx 0.6\,P_{{\rm k}}$ at $r=0.027$~au to $100$~d $\approx 0.6\,P_{{\rm k}}$ at $r=0.6$~au in the case $\beta_0=1$. The oscillation period of the MFT with $\beta_0=0.1$ are smaller and vary form $0.5$~d at $r=0.027$~au to $45$~d at $r=0.6$~au.

Comparison of toroidal magnetic field generation and MFT rise times depicted in Figure~\ref{Fig:Mdot_b}(a) and oscillation periods in Figure~\ref{Fig:Period} shows that $P_{{\rm osc}} \ll t_{{\rm gen}}$. We conclude that the dynamics of the toroidal magnetic field in the considered region proceeds in two stages: slow generation of the toroidal magnetic field ($t_{{\rm gen}}=0.5-10$~yr) with subsequent fast rise and oscillations of the MFT ($P_{{\rm osc}}=1-100$~d).

\subsection{Comparison with observations}
\label{Sec:observ}

As it has been shown in Sections~\ref{Sec:he} and~\ref{Sec:oscill}, thin MFT oscillate near the surface of the disc after rising from interior. Density, radius and temperature periodically change during the oscillations. The oscillations can lead to variability of the accretion disc radiation. MFT contain both gas and refractory dust particles in the considered region. Therefore, the oscillations can cause the IR-variability of the disc radiation. Based on the resistive MHD simulations of the inner region dynamics of the minimum mass solar nebula, \citet{turner10} proposed that magnetic activity can lift dust grains into the disc atmosphere and cause the IR-variability of YSO (see also discussion in \citet{flaherty11}). In this paper, we study similar effect in the frame of slender magnetic flux tube approximation.

IR-variability has been observed in many YSO. \citet{flaherty16} have found that the stars in the Chameleon~I cluster exhibit variability on time-scales of months (20-200 days). Magnitude of the fluctuations ranges from 0.05 to 0.5 mag. In order to test the hypothesis that rising MFT cause the IR-variability of YSO, we perform the simulations of MFT dynamics in two discs of classical T~Tauri stars from the sample presented in \citet{flaherty16}. We have chosen stars J11092266-7634320 and J11100369-7633291 (indexed as 439 and 530, respectively, in Table 1 from \citet{flaherty16}). Masses, accretion rates, luminosities and effective temperatures of these stars are given in columns 2-5 of Table~\ref{Table:C}. Stellar radii are estimated from the relation $L_{\star}=\sigma_{{\rm R}}T_{{\rm eff}}^44\upi R_{\star}^2$ (column 6), where $T_{{\rm eff}}$ is the stellar effective temperature, $R_{\star}$ is the stellar radius. The periods of IR-variations, $\Delta t$, measured by \citet{flaherty16} are listed in column 7. Column~8 gives values of considered $r$-distances from the star. For each star, we simulate the MFT dynamics at two distances: $r_{{\rm i}}$ (values with symbol `i' in brackets in column~8), where temperature is nearly equal to temperature of silicate dust grains evaporation $\sim 1500$~K~\citep[see][]{pollack94}, and $r_{{\rm o}}$ (values with symbol `o' in brackets in column~8), determining the outer boundary of the region of thermal ionization. Radii $r_{{\rm i}}$ and $r_{{\rm o}}$ bound the region of the efficient toroidal magnetic field amplification, where formation of the MFT with both gas and dust is possible. We take magnetic field strength at the stellar surface to be 2 kG for both stars~\citep[see][]{yang11}. Using these stellar parameters, we calculate the structure of the accretion discs using accretion disc model of~\citet{fmfadys}.  Midplane density, temperature, scale height and magnetic field strength of  the discs of stars 439 and 530 are presented in columns 9-12 of Table~\ref{Table:C}.

\begin{table*}
\centering
\caption{Parameters of YSO in Chameleon.}
\label{Table:C}
\begin{tabular}{@{}ccccccccccccc@{}}
\hline 
No & $M_{\star}$ [$M_{\sun}$] & $\dot{M}$ [$M_{\sun}\,{\rm yr}^{-1}$] & $L_{\star}$ [$L_{\sun}$] & $T_{{\rm eff}}$ [K] & $R_{\star}$ [$R_{\sun}$] & $P$ [d] & $r$ [au] & $\rho_{{\rm m}}$ [g cm$^{-3}$] & $T_m$ [K] & $T_{{\rm irr}}$ [K] & $H$ [au] & $B_{{\rm e}}$ [G] \\ 
(1) & (2) & (3) & (4) & (5) & (6) & (7) & (8) & (9) & (10) & (11) & (12) & (13)\\
\hline 
439 & 0.6 & $4.8\times10^{-8}$ & 0.8 & 3669 & 2.2 & 32 & 0.2 (i) & $1.3\times 10^{-8}$ & 1590 & 590 & 0.007 & 18.0 \\ 
 &  &  &  &  &  &  & 0.42 (o) & $2.0\times 10^{-9}$ & 1200 & 410 & 0.019 & 5.0 \\ 
530 & 0.63 & $2\times 10^{-9}$ & 0.66 & 3955 & 1.7 & 35 & 0.07 (i) & $1.6\times 10^{-8}$ & 1460 & 950 & 0.0013 & 20.4 \\ 
 &  &  &  &  &  & & 0.1 (o) & $5.6\times 10^{-9}$ & 1220 & 800 & 0.0024 & 4.0 \\ 
\hline 
\end{tabular} 
\end{table*}

\begin{figure*}
\begin{center}
\includegraphics[width=0.95\textwidth]{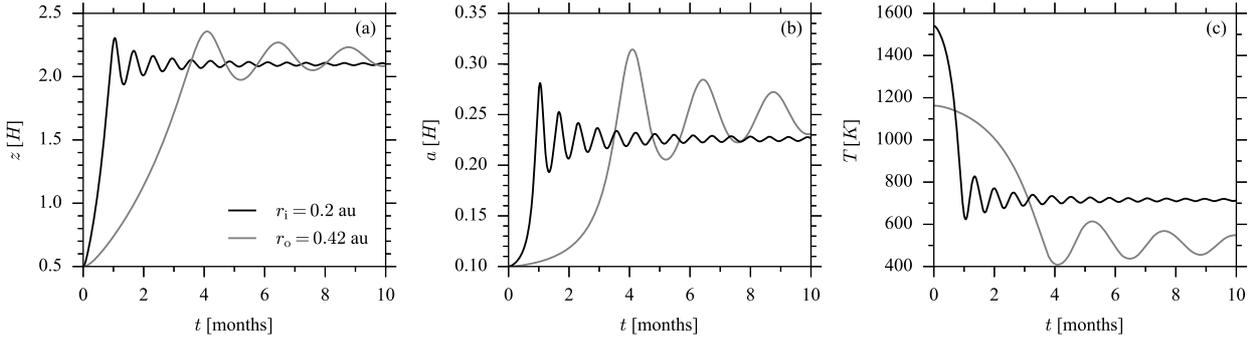}
\end{center}
\caption{The dependences of the coordinate (panel (a)), radius (panel (b)) and temperature (panel (c)) on time for the MFT with $\beta_0=1$, $a_0=0.1\,H$, $z_0=0.5\,H$ in the disc of star 439 in Cha-1. Black line: $r=r_{{\rm i}}$, grey line: $r=r_{{\rm o}}$ (see Table~\ref{Table:C}).}
\label{Fig:Cha1_439}
\end{figure*}

The simulations are carried out for $\beta_0=[0.1, 1]$ and $a_0=0.01\,H\div 0.4\,H$. The simulations show that oscillation periods $P$ weakly depend on the MFT radius. As an example, in Figure~\ref{Fig:Cha1_439} we plot dependences of $z$-coordinate , radius $a$ and temperature $T$ on time for the MFT in the disc of star 439, at $r_{{\rm i}}$ and $r_{{\rm o}}$. Initial MFT parameters: $z_0=0.5\,H$, $a_0=0.1\,H$, $\beta_0=1$. Figure~\ref{Fig:Cha1_439} shows that dynamics of the MFT is similar to that considered in Section~\ref{Sec:oscill}. The MFT rise and then oscillate near the surface of the disc. Oscillation period at $r_{{\rm i}}$
 (20 d) is less than the oscillation period at $r_{{\rm o}}$ (70 d). The MFT experience temperature pulsations around value $T\sim 700$~K at $r_{{\rm i}}$ and around $T\sim 500$~K at $r_{{\rm o}}$. With these temperatures, maximum of emission peaks at wavelengths $\lambda\approx (4-6)\,\mu$m, according to Wien's displacement law.  Maximal temperature variations are $100$~K at $r_{{\rm i}}$ and $200$~K at $r_{{\rm o}}$. 

\begin{figure}
\begin{center}
\includegraphics[width=0.45\textwidth]{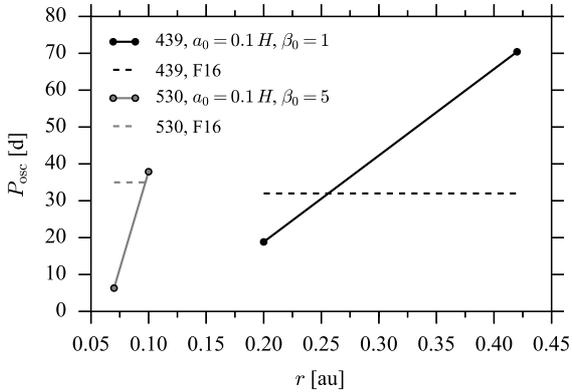}
\end{center}
\caption{The dependence of MFT oscillation periods on the $r$-coordinate in the discs of star 439 (black line), and star 530 (grey line, see Table~\ref{Table:C}).  Black and grey dashed horizontal lines show corresponding observed periods from~\citet{flaherty16}.}
\label{Fig:cha1}
\end{figure}

The dependences of oscillation periods $P_{{\rm osc}}$ on the $r$-distance for each star are shown in Figure~\ref{Fig:cha1}. The results are obtained for $a_0=0.1\,H$, $z_0=0.5\,H$, plasma beta $\beta_0=1$ (star 439) and $\beta_0=5$ (star 530). Star 530 has small accretion rate, and the region of thermal ionization is situated close to the star, $r<0.1$~au. The disc of star 439 has more extended region of thermal ionization, $r<0.42$~au. In both cases, the oscillation periods increase with distance from the star.

For the adopted MFT parameters, the range of oscillation periods is in agreement with the observational values of variability times. Observed period for star 439, $P_{439}=32$~d corresponds to oscillation of the MFT at $r\approx 0.25$~au for $\beta_0=1$. Period of star 530, $P_{530}=35$~d, corresponds to the MFT oscillating at $r=0.1$~au for $\beta_0=5$. Thus, our simulations confirm the hypothesis that rising magnetic fields can be a source of the IR-variability of YSO.

To determine whether the MFT can contribute to IR radiation of considered T~Tauri stars or not, we estimate optical depth of the MFT with respect to IR radiation as $\tau=2a\rho\kappa$, where opacity $\kappa$ is determined as a function of density and temperature according to~\cite{fmfadys}. We get that $\tau$ has the values between 20 and 70 during oscillations at $r=r_{{\rm o}}$ for star 439. Optical thickness for star 530 is of the same order. We conclude that the MFT are very optically thick.

There may be two possible processes causing the variability. First, MFT temperature fluctuations (like in Figure~\ref{Fig:Cha1_439}) can cause the radiation variability. Second, the oscillating MFT can intercept stellar radiation and periodically cast a shadow on the outer disc regions. Stellar radiation reprocessed by the dust in outer disc regions will have an oscillating IR-component.

\section{Conclusion and discussion}
\label{Sec:end}
We investigated numerically the dynamics of slender MFT in the accretion discs of T~Tauri stars. We considered the MFT forming in the region of intense generation of the toroidal magnetic field due to magnetic buoyancy instability. We formulate the equations of slender MFT dynamics taking into account the aerodynamic and turbulent drags.
\cite{kh17raa} have investigated the MFT dynamics taking into account the heat exchange with the external medium of constant temperature. The radiative flux has been estimated in the diffusion approximation. In present work, we investigated additionally influence of the magnetic field of the disc on the MFT dynamics. The vertical structure of the disc was calculated using the polytropic equation of state. In particular, the thermal and magnetic oscillations of the MFT was investigated in this paper.

Simulations were performed for initial radii of the MFT $a_0=0.01\,H\div 0.4\,H$ and radial distances $r=0.027\div 0.6$~au in the accretion disc of solar mass T~Tauri star. Slender flux tube approximation allows us to investigate the dynamics of `soft' MFT with weak magnetic field ($\beta \ga 1$) and `stiff' MFT with strong magnetic field ($\beta<1$). Latter case cannot be considered using multidimensional numerical simulations in frame of classical MHD because of strict limitations on the time step. For comparison, we investigated dynamics of the MFT with  plasma betas $\beta_0=0.01\div 10$. The structure of the accretion disc was calculated with the help of the model of \cite{fmfadys}.

First of all, we considered the dynamics of the MFT in absence of external magnetic field. In this case, we found two regimes of the MFT dynamics. Thin MFT with initial radius $a_0\leq 0.05\,H$ and thick MFT with $a_0\geq 0.16\,H$ rapidly accelerate, then rise with slowly increasing velocity, decelerate a little and acquire nearly steady velocity above the surface of  the disc.
The MFT of intermediate radii $a_0\sim 0.1\,H$ experience thermal oscillations during some time after rising from the disc. The oscillations are due to adiabaticity and slow heat exchange with the  external gas. These oscillations are found only for the MFT with $\beta_0=1$ formed at $r$-distances less than $0.2$~au.  
After radiative heat exchange equalizes internal and external temperatures, the oscillations decay and the MFT continue to move upwards above the disc.

We studied dependence of the MFT velocity, mass and magnetic flux on its initial parameters. Velocity of the MFT increases with initial radius and magnetic field strength. Typical rise velocities are of several km s$^{-1}$. The MFT with weak magnetic field ($\beta_0=10$) have velocities of $0.05-0.5\,{\rm km}\,{\rm s}^{-1}$. The MFT with plasma $\beta=1$ accelerate to velocity $\approx 0.2-4$~km s$^{-1}$ comparable with sound speed, in agreement with analytical estimates of~\cite{kh17raa}. The MFT with strong initial magnetic field (plasma beta $<1$) can reach supersonic velocity up to $\sim 10-15$~km s$^{-1}$. In this case, accurate investigation of MFT supersonic motion with bow shocks can be done by including dependence of the aerodynamic drag on the Mach number. We plan to do this in future works.

The dynamics of the MFT consists in two stages. At the first stage, toroidal magnetic field is generated and MFT form due to magnetic buoyancy instability over time $t_{{\rm gen}}$. At the second stage, the MFT rise from the disc to its suface over the time $t_{{\rm surf}}$. In a process of further upward motion the MFT carry away some mass and magnetic flux to disc atmosphere. Our calculations show that $t_{{\rm surf}}\ll t_{{\rm gen}}$. Therefore, the process of mass and magnetic flux transport from the disk to its atmosphere due to buoyancy is periodic with typical period $t_{{\rm gen}}\sim 0.5-10$~yr. The vertical mass transport rate due to buoyancy $\dot{M}_{{\rm b}}\sim 10^{-12}-10^{-7}\,M_{\sun}\,{\rm yr}^{-1}$. Our calculations show that approximately $20$~\% of disc mass flux can come out from the disc via buoyancy. The rising MFT can be the seed for the formation of jets and outflows from accretion discs. The MFT carry magnetic fluxes $\Phi_{{\rm b}}\sim 10^{19}-10^{24}$~Mx, so that magnetic flux of $\Phi_{{\rm b}}\approx 10^{30}$~Mx, that comprises $20$~\% of total disc magnetic flux, can be carried away from the disc by rising MFT during period of $1$~Myr. We assume that magnetic buoyancy is the mechanism responsible for the efficient magnetic flux escape from the accretion discs of young stars. \cite{kh17ppnl} have shown that formation and rise of the MFT with initial radius $a_0=0.1\,H$ stabilize the strength of the toroidal magnetic field at the level of the poloidal magnetic field strength.

The buoyancy decreases and MFT acquires steady speed above the disc. This is explained by the fact that the external density $\rho_{{\rm e}}$ and the density of MFT decrease with $z$-coordinate. Correspondingly, the buoyant and drag forces also reduce, acceleration of  MFT approaches to zero, MFT moves by inertia and significantly expands above the disc. We interpret this process as a formation of an expanding magnetized `corona' above the disc. 
Expansion of the MFT can be reduced by the internal magnetic tension. We will consider this possibility in future works.

Second important case investigated by us concerns the effect of the magnetic field of the disc on the dynamics of MFT.  In this case,
MFT rise from the disc and start to oscillate at height $z_{{\rm osc}}\approx 2-2.5H\,$ above the midplane of the disc. The magnetic oscillations take place near the point where strengths of internal, $B$,  and external, $B_{{\rm e}}$, magnetic fields are nearly equal. Above this point, the buoyant force is negative since $B<B_{{\rm e}}$ and the MFT has greater density than the external gas. The oscillation periods $P_{{\rm osc}}$ increase with the distance from the star, $P_{{\rm osc}}=(1-10)$ days at $r=0.027$~au and $P_{{\rm osc}}=(1-3)$ months at $r=0.6$~au. The MFT with stronger initial magnetic field (smaller plasma beta) rise to higher altitudes and experience oscillations with smaller periods. The magnetic oscillations decay with time, and occur in the disc over time of the toroidal magnetic field generation. The oscillations of the MFT found in our simulations indicate that the mass carried by the MFT rsing from the disc can come back to the disc. Accumulation of the magnetic flux near the surface of the discs can further lead to the burst release of the magnetic energy as it was suggested by \citet{shibata90}. Magnetic energy of the MFT of $\sim 10^{33}-10^{37}$~erg can be released due to such bursts. Magnetic field $B_{{\rm e}}$ can decrease with height, and the MFT probably can rise to higher altitudes before reaching the point of zero buoyancy in this case.   

Periodic  formation, rise and thermal or magnetic oscillations of the MFT can be interpreted as periodical changes in the accretion disc structure. This process is similar to periodic ejection of buoyant magnetic fields from the disc that has been found in frame of numerical MHD simulations by~\citep[e.g.][]{turner10, takasao18}. At the distance $r=0.6$~au, temperature is $\sim 1000$~K, and the refractory dust grains are present in this region. Therefore, buoyant MFT contain dust, and the oscillations can produce IR-variability of the disc. 

IR-variability is the common feature of YSO (see \citet{flaherty13, flaherty16} and references therein). \citet{flaherty16} have found that the low-mass YSO in the Chameleon I star forming region exhibit variability over time scales of months (20-200 days). Classical T Tauri stars also exhibit variability in optical wavelengths. For example, \cite{rigon17} reported the detection of the optical variability with periods of $20-60$~d of classical T Tauri stars in Taurus-Auriga region.

We simulated MFT dynamics in the discs of two classical T~Tauri stars J11092266-7634320 and J11100369-7633291 (439 and 530) from the sample presented in \citet{flaherty16}. The magnetic oscillation periods found in our simulations are in good agreement with the observed periods for considered stars. Our simulations have shown that variability period of star 439 (32 days) is consistent with the oscillations of the MFT with initial $\beta_0=1$ and $a_0=0.1\,H$. Observed period for star 530 (35 days) can be explained by the oscillations of the MFT with $a_0=0.1\,H$ and $\beta_0=5$. 

Our estimations have shown that the MFT are optically thick with respect to IR-radiation.
The variability can be caused by MFT temperature fluctuations, as well as by periodical screening  the outer disc regions from stellar radiation by the MFT. It should be noted that \cite{takasao18} found formation of the MFT in 3D MHD simulations of the inner regions of the accretion discs and also argued that periodically rising MFT may contribute to the variability of discs radiation.

The MFT can form in the region $r_{{\rm i}}<r<r_{{\rm o}}$~au ($r_{{\rm i}}=r(T=1500\,K)$, $r_{{\rm o}}$ is the outer boundary of the zone of the thermal ionization), where magnetic field is efficiently amplified and temperature is lower than 1500~K, so that the MFT contain the refractory dust particles. For stars 439 and 530 radii $r_{{\rm i}}=0.2$~au, $0.07$~au and $r_{{\rm o}}=0.42$~au, $0.1$~au, respectively. Therefore, the oscillations occur in a range of radii, which is consistent with conclusions of~\cite{flaherty16}. Magnetic oscillations decay with time. For example, magnitude of temperature oscillations reduces by $100$~K over $1.3$~months at $r_{{\rm i}}=0.2$ and over $4.7$~months at $r_{{\rm o}}=0.42$ in the disc of star J11092266-7634320.  Thus, significant temperature variations of oscillating MFT, $\Delta T \sim 100$~K,  can be detected within  time of $200$~d corresponding to observation time in~\citet{flaherty16}. 

\citet{tambov08} suggested that inhomogeneities in the disc centrifugal winds containing both gas and dust can cause variations in circumstellar extinction observed in T~Tauri stars. Our simulations show that magnetic buoyancy can transport mass from the accretion disc to its atmosphere. We propose that such this process can be responsible for the variations in circumstellar extinction observed in T~Tauri stars. Similar hypothesis was proposed by~\citet{miyake16}, who simulated the time evolution of the dust grain distribution in the vertical direction inside the minimum mass solar nebula.

As \cite{schram96} have shown, longitudinal perturbations of the magnetic flux tubes can lead to the formation of the arcs above the disc. We assume that such orbiting arcs can cast shadows on the outer regions of the accretion discs. This problem needs further investigation.

Toroidal magnetic field in the disc is generated from the poloidal one over the time scale of rotation period. Tension of the poloidal magnetic field lines can slow down the disc rotation, and, as a consequence, hinder the generation of the toroidal magnetic field in the considered region. This effect takes place if the energy of the poloidal magnetic field is comparable with the rotational energy.  This condition for the keplerian disc can be written as
\begin{equation}
	\beta \sim 2\times 10^{-4}\left(\frac{H/r}{0.01}\right)^2,
\end{equation}
i.e. the magnetic field with $\beta \sim 10^{-4}$ will influence significantly the disc rotation. Kinematic approximation $\beta>1$ is used in our model of the accretion disc, and the poloidal magnetic field cannot be strong enough to stop generation of the toroidal magnetic field and formation of the MFT.

We did not consider the effect of the azimuthal velocity shear on the MFT dynamics. \citet{shibata90} have found that rise velocity of the MFT in the presence of velocity shear is smaller than in the case of no shear. MHD simulations of MFT rising from  the upper convection zone of the Sun to the solar atmosphere have shown that the MFT dynamics is sensitive to the twist of the MFT~\citep[see][]{fan98, magara01, sykora15}. Strongly twisted MFT retain their coherent structures during the rise, while the MFT without twist splits into a vortex pair and lose significant amount of their magnetic flux. Up to date, there are no detailed simulations of the MFT formation in the accretion discs. It is hard to make conclusions about degree of their twist. In our simulations we implicitly assumed that the MFT retain their coherence and do not lose the magnetic flux. Probably, loss of the magnetic flux would lead to decrease of the MFT velocity.

The magnetic pressure acts only in the direction perpendicular to the magnetic lines. Uniform magnetic field of the disc $B_{{\rm e}}$ considered in this work will lead to flattening of the MFT in the direction of motion. Generally speaking, the toroidal magnetic field is generated not only in the region of MFT formation, but in the entire volume of the disc. Therefore, toroidal magnetic field of the disc will influence MFT dynamics together with the poloidal magnetic field. Assuming that intensities of the toroidal and poloidal magnetic fields are comparable, we adopted  in this work that the magnetic pressure is isotropic, and the cross-section of the MFT remains nearly round. Effects of non-uniform MFT expansion can be investigated in two-dimensional model of the MFT dynamics, that we aim to elaborate in future.
We plan to develop more detailed model of the vertical structure of the disc in the future. Interesting task is investigation of the dynamics of the magnetic rings in the accretion discs of young stars, i.e. study the evolution of major radius of the toroidal MFT. 

\section*{Acknowledgements}
The work of AED is supported by the Russian Foundation for Basic Research (project 18-02-01067). The work of SAKh and AMS is supported by the Ministry of Science and High Education (the basic part of the State assignment, RK No. AAAA-A17-117030310283-7) and by the Act 211 Government of the Russian Federation, contract No. 02.A03.21.0006. The authors thank prof. Dmitry Bisikalo, Dr.~Sergey Parfenov, Dr.~Anna Evgrafova and Dr. Vitaly Akimkin for useful comments. We are also grateful to Lyudmila Lapina for checking the English language in the paper. We thank anonymous referee for a detailed review and useful comments.

\bibliographystyle{mnras}
\bibliography{dudorov}

\bsp	
\label{lastpage}

\end{document}